\documentclass[PAPER, atlasdraft=false, cernpreprint, USenglish, texmf, orcidlogo]{atlasdoc}
\usepackage{atlaspackage}
\usepackage{atlasbiblatex}

\usepackage{float}
\usepackage{siunitx}
\usepackage{booktabs}
\usepackage{pdflscape}
\usepackage{subfig}
\usepackage{multirow}

\usepackage{atlasphysics}
\graphicspath{{logos/}{figures/}}

\usepackage{ANA-TOPQ-2019-18-PAPER-defs}

\AtlasTitle{A search for electroweak $t\bar{t}W\kern-0.2em j$ production in multileptonic final states at $\sqrt{s} = 13$~TeV with the ATLAS detector and bounds on effective field theory operators}

\AtlasAbstract{%
A search is presented for the electroweak production of a top quark pair in association with a $W$ boson and at least one additional jet, known as the ${t\bar{t}W\kern-0.2em j_{\text{EW}}}$ process. This process has embedded within it a $tW$-scattering vertex, which is probed directly for the first time. The collision data were collected with the ATLAS detector during Run 2 of the LHC and correspond to an integrated luminosity of 140 $\mathrm{fb^{-1}}$ at $\sqrt{s}=13$~TeV. The search uses same-charge pairs of electrons and muons together with jets, of which at least one is $b$-tagged. The properties of the most forward jet relative to the rest of the event are used to discriminate the electroweak ${t\bar{t}W\kern-0.2em j_{\text{EW}}}$ production process from its strong $t\bar{t}W$ production counterpart. A measured (expected) 95\% CL upper limit on the cross section is set at $\sigma_{{t\bar{t}W\kern-0.2em j_{\text{EW}}}} < 251$~fb (230~fb), to be compared with the expected SM cross section of 47.7~fb. Limits are set on the SM Effective Field Theory (EFT) operators $O_{Ht}$ and $O^{(1)}_{HQ}$, which modify the electroweak couplings of the top quark through contributions to the $tW$-scattering vertex. The interpretation acts as a case study to emphasize the importance of energy-dependent sensitivity, multi-process and multi-operator EFT contributions.}

\AtlasRefCode{TOPQ-2019-18}

\PreprintIdNumber{CERN-EP-2025-207}

\AtlasJournal{Phys.\ Rev.\ D.}
\hypersetup{pdftitle={ATLAS document},pdfauthor={The ATLAS Collaboration}}

\begin{document}

\maketitle

\setcounter{tocdepth}{1}
\tableofcontents

\clearpage
\section{Introduction}

\begin{figure}[b!]
\centering
\subfloat[]{\includegraphics[width=0.24\textwidth]{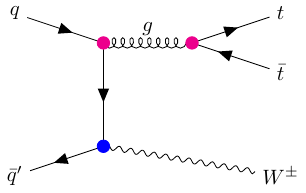}\label{fig:ttw-lo}}
\subfloat[]{\includegraphics[width=0.24\textwidth]{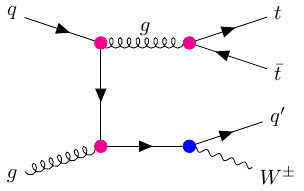}\label{fig:ttw-nlo1}}
\subfloat[]{\includegraphics[width=0.24\textwidth]{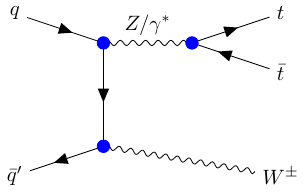}\label{fig:ttw-lo3}}
\subfloat[]{\includegraphics[width=0.24\textwidth]{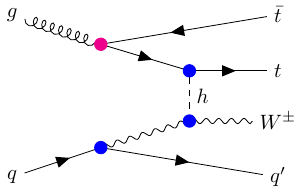}\label{fig:ttw-nlo3}}
\caption[dummy]{
Illustrative Feynman diagrams for the dominant production modes of \ttW~\cite{TOPQ-2019-30}: \subref{fig:ttw-lo} the leading-order (LO) contribution ($\alpha\alphas^2$), \subref{fig:ttw-nlo1} a real emission diagram from the NLO QCD contribution ($\alpha\alphas^3$), \subref{fig:ttw-lo3} the tree-level EW contribution ($\alpha^3$), and \subref{fig:ttw-nlo3} a representative diagram of the combined NLO QCD and EW contributions ($\alpha^3\alphas$). The pink circles correspond to QCD couplings and the blue circles correspond to EW couplings. }
\label{fig:feynmandiagrams}
\end{figure}

One of the key physics goals of the Large Hadron Collider (LHC) physics program is to probe the nature of electroweak (EW) symmetry breaking (EWSB), both in the Standard Model (SM) and beyond (BSM). The production of top quarks in association with EW bosons offers a unique testing ground to answer these questions. In particular, \ttW production, which superficially appears to be a simple process composed of well-known SM couplings, proves to have a rich phenomenology. The EW production of \ttW in association with an additional jet, \ttWjEW, is especially interesting because it probes new dynamics at the LHC, the \tW-scattering vertex.
The dominant contributing Feynman diagrams for the \ttW process, including
additional quantum chromodynamics (QCD) and EW corrections, are presented in
Figure~\ref{fig:feynmandiagrams}, where
Figure~\ref{fig:ttw-nlo3} shows an example of one \ttWjEW
contribution. The \ttWjEW process produces a `forward' jet with greater pseudorapidity than the QCD \ttW production process. This forward jet can be used to separate the processes experimentally.
The \tW-scattering vertex provides strong energy-enhanced sensitivity potential to BSM effects not present in the SM EWSB\@. These EWSB-specific BSM signatures can be investigated through effective field theory (EFT) approaches. A deviation of the top EW couplings from the SM with a signature of energy growth that diverges quadratically with momentum would be a signal of BSM physics. This would indicate that the top quark is strongly coupled to the sector responsible for EWSB~\cite{Dror:2015nkp,Maltoni:2019aot}.

The \ttWjEW process can be understood in the wider context of
next-to-leading-order (NLO) QCD and EW corrections to \ttW
production. The traditionally `leading' \NLO EW corrections (of order
$\alpha^2\alphas^2$, where $\alpha$ and $\alphas$ refer to the
electroweak and strong coupling constants, respectively) cause
few-percent negative changes to the inclusive \ttW cross section;
however, even larger positive contributions from traditionally
`sub-leading' \NLO EW corrections (of order $\alpha^3\alpha_s$) were
pointed out in
Refs.~\cite{Dror:2015nkp,Frederix:2017wme,FebresCordero:2021kcc}. It is
this sub-leading contribution, of order $\alpha^3\alpha_s$, which
corresponds to the \ttWjEW signal considered here.

Measurements of \ttWjEW production also provide a better understanding of \ttW production. The \ttW process is an important background in LHC searches and precision measurements, as one of the few SM processes that provides an irreducible source of same-charge dilepton and trilepton events. This is especially important given existing observations of tension with the SM predictions in direct measurements of \ttW production. Measurements of \ttW production by the ATLAS and CMS Collaborations at $\sqrt{s}=8$~\TeV~\cite{TOPQ-2013-05,CMS-TOP-14-021} were in agreement with the SM\@. However, at $\sqrt{s}=13$~\TeV higher precision measurements~\cite{TOPQ-2019-30,TOPQ-2015-22,TOPQ-2016-11,CMS-TOP-17-005,CMS-TOP-21-011}
observe tensions with the SM predictions. These tensions are somewhat alleviated by the inclusion of higher-order effects~\cite{Frederix:2017wme,Frederix:2021agh, PhysRevLett.131.231901}. Therefore, tests of the modeling of higher-order effects, i.e., \ttW with additional jets, are especially important for measurements where this is the dominant background, such as \ttH~\cite{HIGG-2017-02,CMS-HIG-19-008}
and \tttt~\cite{TOPQ-2021-08,CMS-TOP-22-013}.

\subsection{EFT sensitivity}

\begin{figure}[b!]
\begin{center}
\includegraphics[scale=0.9,trim=0 0 1cm 0, clip]{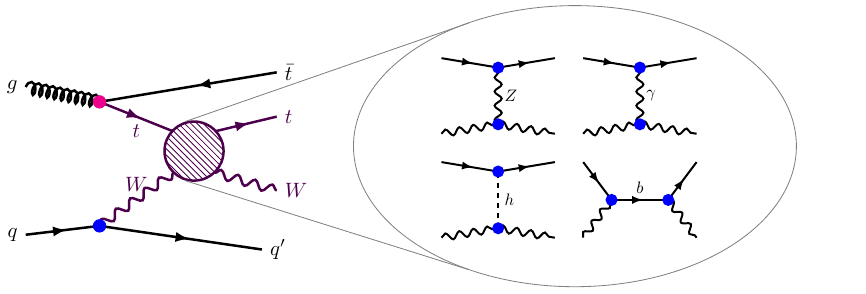}
\caption{A schematic representation of the $tW$-scattering vertex in the Feynman diagram of the \ttWjEW process, showing different contributing sub-processes, modified from Ref.~\cite{Dror:2015nkp}.}
\label{fd}
\end{center}
\end{figure}

As shown in Figure~\ref{fd}, the \ttWjEW process can be considered to contain an effective vertex probing various SM couplings such as the top-quark $Z$ boson ($t$--$Z$) and top Yukawa ($y_t$) couplings.
The sensitivity of this vertex to new physics has a significant quadratic energy dependence due to the scattering topology, contrasting with at most linear energy dependence in \ttZ and \ttH processes. This enhanced energy dependence is also true of BSM EWSB EFT operators that can contribute to \ttWjEW production.

This unique sensitivity to the $t$--$Z$ coupling, which is complementary to dedicated measurements, serves as an additional handle on deviations from the SM that could be a sign of new physics effects in the EWSB mechanism~\cite{Aguilar-Saavedra:2009xmz,Aguilar-Saavedra:2013qpa,Perelstein:2005ka,Chivukula:1992ap,Chivukula:1994mn,PhysRevD.52.5374,PhysRevD.55.5848,PhysRevD.56.402}. The different structure of \ttWjEW production, compared with \ttZ production, lifts degeneracies when considering multiple EFT operators, such as \Opt and \OHQone.

The $tW$-scattering vertex shown in Figure~\ref{fd} is also sensitive to EFT operators affecting the top-quark Higgs boson coupling, such as $\ensuremath{O_{tH}}\xspace$. However, the sensitivity is less than in dedicated measurements~\cite{HIGG-2022-17} and therefore these are not considered in this analysis.

EFT frameworks are a consistent and model-independent way to search for new physics at high energies.
A general SM EFT (SMEFT) Lagrangian considering operators up to mass-dimension-six takes the form
\begin{equation*}
\mathcal{L}_{EFT} = \mathcal{L}_{SM}  + \sum_{i} \frac{c_i}{\Lambda^2} \mathcal{O}_{i},
\end{equation*}
where $\mathcal{L}_{SM}$ is the SM Lagrangian, the $\mathcal{O}_{i}$
are the operators of mass-dimension-six and the $c_i$ are the
associated Wilson coefficients for a given general cut-off scale $\Lambda$. In a measurement only the ratio $c_i/\Lambda^2$ is relevant, and here $\Lambda = 1$~\TeV is used to set the counting scale of $c_i$. The cut-off scale for the EFT corresponds to the energy scale where new physics would arise. Using the notation defined in Ref.~\cite{eftLHC}, the two operators of relevance for this analysis are

\begin{equation*}
\frac{c_{Ht}}{\Lambda^2} \Opt = \frac{c_{Ht}}{\Lambda^2} ( H^{\dagger} \overset{\leftrightarrow}{iD_{\mu}} H ) (\bar{t}_R \gamma^{\mu} t_R)
\end{equation*}
and
\begin{equation*}
\frac{c^{(1)}_{HQ}}{\Lambda^2} \OHQone = \frac{c^{(1)}_{HQ}}{\Lambda^2} ( H^{\dagger} \overset{\leftrightarrow}{D_{\mu}} H ) (\bar{Q}_L \gamma^{\mu} Q_L).
\end{equation*}

These operators have a strong effect on the \ttWjEW process at current experimental constraints. In \ttZ production EFT effects for these operators manifest mainly through corrections that are linear in $c_i/\Lambda^2$, whereas in \ttWjEW the energy growth leads to a significant quadratic dependence in $c_i/\Lambda^2$ where the linear interference term is strongly suppressed due to symmetries in the SM amplitude~\cite{Maltoni:2019aot}. This phenomenology arises because of the presence of two longitudinally polarized gauge bosons in the \tW-scattering topology but not in \ttZ production~\cite{Dror:2015nkp,Maltoni:2019aot}. The EFT can still be valid in the perturbative expansion in this case even when the linear term is very small compared with the quadratic term~\cite{PhysRevD.95.065014}. Several EFT interpretation configurations are considered in this measurement and it is shown that additional constraints from \ttWjEW production can provide separation between the \Opt and \OHQone operators, which can otherwise suffer from degenerate directions in the measurement.

\subsection{Analysis strategy}

A signal-enriched region (SR) is defined to target the signal \ttWjEW production process by selecting events that contain two same-charge leptons (electrons or muons) and at least four jets. At least one jet is required to be identified as containing a $b$-hadron ($b$-tagged). In addition, control regions (CRs) are defined that are enriched in and used to normalize background processes.

The signal contribution in the SR is then measured with a simultaneous binned profile-likelihood fit exploiting the difference in kinematics between the signal and the dominant QCD-induced $\ttW{}j$ background.

Finally, an EFT interpretation is performed, using Monte Carlo (MC) samples to model the contributions of the \Opt and \OHQone operators on both the \ttWjEW and \ttZ processes. Sensitivity to \ttZ comes primarily from the CR enriched in this process. The EFT operators are considered both independently and simultaneously and $95\%$ confidence intervals are determined.

The paper is structured as follows: Section~\ref{sec:Detector} briefly describes the ATLAS detector, Section~\ref{sec:DataMC} provides an overview of the data and MC samples used in the measurements, and Section~\ref{sec:objects} details the object reconstruction and selection. Section~\ref{sec:EventSelection} describes the event selection criteria, and data-driven background estimates are described in Section~\ref{sec:NPFakes}. An overview of the systematic uncertainties is given in  Section~\ref{sec:Systematics}, the results of the SM measurement are presented in Section~\ref{sec:StatisticalAnalysis}, and the EFT interpretation is detailed in Section~\ref{sec:EFTInterpretation}. Finally, conclusions are given in Section~\ref{sec:Conclusions}.

\clearpage

%


\newcommand{\AtlasCoordFootnote}{%
ATLAS uses a right-handed coordinate system with its origin at the nominal interaction point (IP)
in the center of the detector and the \(z\)-axis along the beam pipe.
The \(x\)-axis points from the IP to the center of the LHC ring,
and the \(y\)-axis points upwards.
Polar coordinates \((r,\phi)\) are used in the transverse plane,
\(\phi\) being the azimuthal angle around the \(z\)-axis.
The pseudorapidity is defined in terms of the polar angle \(\theta\) as \(\eta = -\ln \tan(\theta/2)\) and is equal to the rapidity
$ y = \frac{1}{2} \ln \left( \frac{E + p_z}{E - p_z} \right) $ in the relativistic limit.
Angular distance is measured in units of \(\Delta R \equiv \sqrt{(\Delta y)^{2} + (\Delta\phi)^{2}}\).}

\section{ATLAS detector}
\label{sec:Detector}

The ATLAS detector~\cite{PERF-2007-01} at the LHC covers nearly the entire solid angle around the collision point.\footnote{\AtlasCoordFootnote}
It consists of an inner tracking detector surrounded by a thin superconducting solenoid, electromagnetic and hadronic calorimeters,
and a muon spectrometer incorporating three large superconducting air-core toroidal magnets.

The inner-detector system (ID) is immersed in a \qty{2}{\tesla} axial magnetic field
and provides charged-particle tracking in the range of \(|\eta| < 2.5\).
The high-granularity silicon pixel detector covers the vertex region and typically provides four measurements per track,
the first hit generally being in the insertable B-layer (IBL) installed before Run~2~\cite{ATLAS-TDR-19,PIX-2018-001}.
It is followed by the SemiConductor Tracker (SCT), which usually provides eight measurements per track.
These silicon detectors are complemented by the transition radiation tracker (TRT),
which enables radially extended track reconstruction up to \(|\eta| = 2.0\).
The TRT also provides electron identification information
based on the fraction of hits (typically 30 in total) above a higher energy-deposit threshold corresponding to transition radiation.

The calorimeter system covers the pseudorapidity range \(|\eta| < 4.9\).
Within the region \(|\eta|< 3.2\), electromagnetic calorimetry is provided by barrel and
endcap high-granularity lead/liquid-argon (LAr) calorimeters,
with an additional thin LAr presampler covering \(|\eta| < 1.8\)
to correct for energy loss in material upstream of the calorimeters.
Hadronic calorimetry is provided by the steel/scintillator-tile calorimeter,
segmented into three barrel structures within \(|\eta| < 1.7\), and two copper/LAr hadronic endcap calorimeters.
The solid angle coverage is completed with forward copper/LAr and tungsten/LAr calorimeter modules
optimized for electromagnetic and hadronic energy measurements, respectively.

The muon spectrometer (MS) comprises separate trigger and
high-precision tracking chambers measuring the deflection of muons in a magnetic field generated by the superconducting air-core toroidal magnets.
The field integral of the toroids ranges between \num{2.0} and \qty{6.0}{\tesla\metre}
across most of the detector.
Three layers of precision chambers, each consisting of layers of monitored drift tubes, cover the region \(|\eta| < 2.7\),
complemented by cathode-strip chambers in the forward region, where the background is highest.
The muon trigger system covers the range \(|\eta| < 2.4\) with resistive-plate chambers in the barrel, and thin-gap chambers in the endcap regions.

The luminosity is measured mainly by the LUCID--2~\cite{LUCID2} detector that records Cherenkov light produced in the quartz windows of photomultipliers located close to the beampipe.

Events were selected by the first-level trigger system implemented in custom hardware,
followed by selections made by algorithms implemented in software in the high-level trigger~\cite{TRIG-2016-01}.
The first-level trigger accepted events from the \qty{40}{\MHz} bunch crossings at a rate close to \qty{100}{\kHz},
which the high-level trigger further reduced to record complete events to disk at about \qty{1.25}{\kHz}.

A software suite~\cite{SOFT-2022-02} is used in data simulation, in the reconstruction
and analysis of real and simulated data, in detector operations, and in the trigger and data acquisition
systems of the experiment.


%
%
%

\section{Data and simulated event samples}
\label{sec:DataMC}

The data used is the full LHC \RunTwo (2015--2018) recorded by the ATLAS detector at $\sqrt{s} = 13$~\TeV, which corresponds to an integrated luminosity of 140~$\mathrm{fb}^{-1}$.
Only events from time periods in which LHC beams were stable and all ATLAS detectors were operational are selected~\cite{DAPR-2018-01}.
The data are collected using a combination of single-electron and single-muon triggers, with requirements on the identification, isolation, and transverse momentum (\pT) of the leptons to maintain high efficiency across the full momentum range while controlling the trigger rates~\cite{TRIG-2016-01,TRIG-2018-05,TRIG-2018-01}. For electrons the trigger thresholds for 2016--2018 (2015) data are \pT = 26, 60 and 140~\GeV (\pT = 24, 60 and 120~\GeV), whereas for muons the thresholds are \pT = 26 and 50~\GeV (\pT = 20 and 50~\GeV). Isolation requirements are applied to the triggers with the lowest \pT threshold for both electrons and muons~\cite{ATL-DAQ-PUB-2016-001,ATL-DAQ-PUB-2017-001, ATL-DAQ-PUB-2018-002, ATL-DAQ-PUB-2019-001}.

Samples of simulated events are used to model the signal and the prompt SM background. The samples are processed to include the effects of the ATLAS detector with a simulation in \GEANT~\cite{Agostinelli:2002hh}. All simulated events are processed with the same reconstruction software as the data and corrected so that the object reconstruction, identification and trigger efficiencies, energy scales, and energy resolutions match the data. The effect of multiple interactions in the same and neighboring bunch crossings (pileup) is modeled by overlaying each simulated hard-scattering event with inelastic $pp$ events generated with \PYTHIA[8.1]~\cite{Sjostrand:2007gs} using the NNPDF2.3LO~\cite{Ball:2012cx} set of parton distribution functions (PDF) and the A3 set of tuned parameters. Separate MC production campaigns are used to model the different pileup distributions observed in the 2015--2016, 2017 and 2018 data. The samples of simulated events are reweighted to reproduce the observed distribution of the average number of collisions per bunch crossing in each data-taking period.

In all cases, the top quark mass is set to 172.5~\GeV. Unless specified below, samples interfaced to \PYTHIA[8.2]~\cite{Sjostrand:2014zea} use NNPDF3.0NLO~\cite{Ball:2014uwa} and NNPDF2.3LO PDF sets for the matrix element (ME) and the parton showering respectively, and the A14 tune. The decays of bottom and charm hadrons are simulated using the \EVTGEN[1.2-1.6] program~\cite{Lange:2001uf} for all generators except \Sherpa.

The \ttWjEW signal sample is generated with \SHERPA[2.2.10]~\cite{Bothmann:2019yzt,Gleisberg:2008ta} at leading-order (LO), namely $\mathcal{O} (\alpha_s \alpha^3)$, for the ME and showered with the default \SHERPA[2.2.10] settings and NNPDF3.0NNLO~\cite{NNPDF:2017mvq}.
The renormalization and factorization scales, $\mu_\mathrm{R}$ and $\mu_\mathrm{F}$, are set to $H_\textrm{T}/2 = \frac{1}{2}\sum_i\sqrt{m_i^2 + p_{\text{T},i}^2}$ where $i$ runs over all outgoing partons from the ME calculation.
The cross section from the generator is $\sigma_{\ttWjEW} = 42.1^{+1.7}_{-1.3}$~fb. This is scaled (by a factor of 1.13) to match the theoretical cross section predicted at \NLO in QCD from Ref.~\cite{Frederix:2021agh}, $\sigma_{\ttWjEW} = 47.7^{+1.9}_{-1.4}$~fb.

The \ttW QCD sample is generated with \SHERPA[2.2.10] at \NLO accuracy in QCD with $\mu_R$ = $\mu_F$ = $H_\textrm{T}/2$. The ME was calculated for up to one additional parton at \NLO and up to two partons at \LO using
\textsc{Comix}~\cite{Gleisberg:2008fv}
and \OPENLOOPS~\cite{Cascioli:2011va,Denner:2016kdg,Buccioni:2019sur}, and merged with the \Sherpa
parton shower~\cite{Schumann:2007mg} using the \MEPSatNLO
prescription~\cite{Hoeche:2012yf,Hoeche:2011fd,Hoeche:2012yf,Catani:2001cc,Hoeche:2009rj} with a merging scale of 30 \GeV. The sample is normalized to the next-to-next-to-leading-order (\NNLO) cross section of $710.7^{+37.4}_{-48.1}$~fb quoted in Ref.~\cite{PhysRevLett.131.231901}.

The $\ttbar Z/\gamma^*$, $\ttbar H$, $tqZ/\gamma^*$, $tqH$, $tWZ$ and $tWH$ samples are generated at \NLO with \MGNLO and showered with \PYTHIA[8.2]. These samples are normalized to the best theoretical predictions at \NLO using k-factors ~\cite{deFlorian:2016spz}. Other minor backgrounds with top quarks ($t\bar{t}t$, $t\bar{t}t\bar{t}$, $t\bar{t}W^+W^-$) are generated at \LO with \MGNLO and showered with \PYTHIA[8.2].

Diboson ($VV$) processes with four charged leptons, three charged leptons and a neutrino, or two charged leptons and two neutrinos (all with or without extra partons) are simulated with \SHERPA[2.2.2] with the PDF set NNPDF3.0NNLO\@. The processes are generated with up to one extra parton at \NLO and up to three extra partons at \LO using the same formalism as in \ttW.

Non-prompt and fake leptons can be generated from many different sources. Non-prompt leptons originate mainly from semileptonic hadron decays, while fake leptons are mostly due to misidentified jets. These two categories are jointly hereafter referred to as `fake leptons'. The fake-lepton background, primarily arising from \ttbar production, is derived from a data-driven approach. The MC production of \ttbar events is modeled using the \POWHEGBOX[2] generator~\cite{Alioli:2010xd} at \NLO with the \NNPDF[3.0nlo] PDF set and the damping factor $h_\text{damp}$ set to 1.5 times the top quark mass. The events are interfaced with \PYTHIA[8.2].
Single top-quark production in the $s$- and $t$-channels, and $tW$ production, is simulated similarly.

The production of \Wboson and \Zboson bosons with multiple jets ($V+$jets) is simulated with the \SHERPA[2.2.1] generator using NLO-accurate MEs for up to two jets and LO-accurate MEs for up to four jets, calculated with the Comix~\cite{Gleisberg:2008fv} and \OPENLOOPS libraries. They are matched with the \SHERPA parton shower using the \MEPSatNLO prescription.
The \NNPDF[3.0nnlo] set of PDFs is used and the samples are normalized to \NNLO predictions.

Processes with three heavy gauge bosons ($WWW$, $WWZ$, $WZZ$ and $ZZZ$) with up to six leptons in the final state are simulated with \SHERPA[2.2.2] and the \NNPDF[3.0nlo] PDF set.
Final states with no additional partons are calculated at NLO, whereas final states with one, two or three additional partons are calculated at LO.

For the dominant prompt processes, alternative MC samples are used to estimate modeling uncertainties. These are described in more detail in Section~\ref{sec:Systematics}.

Additional MC samples for EFT contributions are produced at LO in QCD for the \ttWjEW and \ttZ processes, using the \MADGRAPH[v2.9.3] generator and \PYTHIA[8.2] parton shower. The \texttt{SMEFTsim 3.0} model~\cite{Brivio:2020onw} is used as the Universal FeynRules Output (UFO)~\cite{Degrande:2011ua}, in the $M_W$ electroweak input scheme~\cite{Brivio:2021yjb} with the \texttt{top} flavor restrictions.
The nominal events are generated according to the SM, and the \MADGRAPH reweighting module is used to compute alternative event weights corresponding to the inclusion of dimension-6 EFT vertices and linearized propagator corrections in the production Feynman diagrams.
The event weights are used to extract the dependence of various observables, including the cross section, on different EFT operators. The EFT samples are processed to include the effects of the ATLAS detector with a fast simulation in \textsc{AtlFast~II}~\cite{SOFT-2010-01} and are reconstructed to perform the EFT interpretation in Section~\ref{sec:EFTInterpretation}.


%

\section{Object reconstruction}
\label{sec:objects}

Interaction vertices from the $pp$ collisions are reconstructed from at least two tracks with \pt larger than $500~\mev$ that are consistent with originating from the beam collision region in the $x$--$y$ plane. If more than one primary vertex candidate is found in the event, the candidate for which the associated tracks form the largest sum of squared \pt is selected as the hard-scatter primary vertex~\cite{ATL-PHYS-PUB-2015-026}.

Electron candidates are reconstructed from energy clusters in the electromagnetic calorimeter matched to a track in the ID~\cite{EGAM-2018-01}.
They are required to satisfy $\textrm{\pT} > 7$~\GeV and $|\eta|<2.47$, excluding the transition region between the endcap and barrel calorimeters ($1.37<|\eta|<1.52$).
`Loose' and `tight' electron identification working points are used, which are based on a likelihood discriminant employing calorimeter, tracking and combined variables that provide separation between electrons and jets. Electrons are further required to satisfy a boosted decision tree (BDT) discriminant threshold designed to reject charge misidentification.

Muon candidates are reconstructed by performing a combined fit to the track information from the ID and MS~\cite{MUON-2018-03}. The resulting muon candidates are re-fitted using the complete track information from both detector systems. They are required to satisfy $\textrm{\pT} > 7$~\GeV and $|\eta| < 2.5$. The `medium' muon identification working point is used.

Electron (muon) candidates are matched to the primary vertex by
requiring that the significance of their transverse impact parameter,
$d_0$,\footnote{The transverse impact parameter, $d_0$, is
defined in the $x$--$y$~plane as the distance of closest approach of the
track to the beamline.} satisfies $|d_0/\sigma(d_0)|<5\,(3)$, where
$\sigma(d_0)$ is the measured uncertainty in $d_0$, and by requiring
that their longitudinal impact parameter, $z_0$,\footnote{The longitudinal impact parameter, $z_0$, is defined as the distance in $z$ between the primary vertex and the point on the track used to evaluate $d_0$.} satisfies $|z_0 \sin\theta|<0.5$~mm.

Leptons from heavy-flavor hadron decays, misidentified jets, or photon conversions are suppressed using a BDT discriminant referred to as the non-prompt lepton BDT~\cite{MUON-2018-03}. This BDT uses isolation and lifetime information associated with a track jet that matches the electron or muon candidate.  A `tight' working point is defined with a prompt electron (muon) efficiency of $65\%$ ($80\%$) for a candidate lepton $\pt=20$~\GeV, increasing to $90\%$ ($95\%$) at $\pt=45$~\GeV. The rejection factors for non-prompt electrons (muons) from heavy-flavor decays range from 60 (30) at low \pt to 30 (20) at high \pt. The rejection factor for electrons from photon conversions is $5$ at low \pt and $1.5$ at high \pt.

The constituents for jet reconstruction are identified by combining measurements from both the ID and the calorimeter using a particle flow (PFlow) algorithm~\cite{PERF-2015-09}. Jet candidates are reconstructed from `PFlow objects' using the \antikt algorithm~\cite{Cacciari:2008gp,Fastjet} with a radius parameter of $R=0.4$. They are calibrated using simulation with corrections obtained by using in situ techniques in data~\cite{JETM-2018-05}. Only jet candidates with $\pT> 25$~\GeV\ and within $|\eta|<4.5$ are selected. To reduce the effect of \pileup, each jet with $\pT<60$~\GeV\ and $|\eta|<2.4$ must satisfy the `tight' working point of the jet vertex tagger (JVT)~\cite{PERF-2014-03} used to identify jets that originate from the primary vertex. A set of quality criteria is also applied to reject events containing at least one jet arising from non-collision sources or detector noise~\cite{ATLAS-CONF-2015-029}.

Jets containing $b$-hadrons are identified ($b$-tagged) with an algorithm~\cite{FTAG-2019-07} that uses a deep-learning neural network based on the distinctive features of $b$-hadron decays, primarily the impact parameters of tracks and the displaced vertices reconstructed in the ID\@. Additional input to this network is provided by discriminating variables constructed by recurrent neutral networks~\cite{ATL-PHYS-PUB-2017-003}, which exploit the spatial and kinematic correlations between tracks originating from the same $b$-hadron. A multivariate $b$-tagging discriminant value is calculated for each jet. A jet is considered $b$-tagged if it satisfies $|\eta|<2.4$ and passes the working point corresponding to $77\%$ average expected efficiency to tag a $b$-quark jet, with rejection factors of $5.6$ and $192$ for $c$-jets and light-flavor jets,\footnote{`Light jet' refers to a jet originating from the hadronization of a light quark ($u$, $d$, $s$) or a gluon.} respectively.
Correction factors derived from dedicated calibration samples enriched in \bjet{}s, $c$-tagged jets, or light-tagged jets are applied to the simulated event samples~\cite{FTAG-2018-01,FTAG-2020-08,FTAG-2019-02}.

To avoid double-counting of detector signatures, objects are removed using the angular distance \(\Delta R \equiv \sqrt{(\Delta y)^{2} + (\Delta\phi)^{2}}\), where $y$ is the rapidity. The overlap removal is performed in the following order: electrons sharing a track with a muon; jets within $\Delta R=0.2$ of an electron; electrons within $\Delta R=0.4$ of a jet; jets within $\Delta R=0.4$ of a muon if they have at most two associated tracks; muons within $\Delta R=0.4$ of a jet.

The missing transverse momentum $\vec{p}_{\mathrm{T}}^{\mathrm{miss}}$ (with magnitude $\met$) is defined as the negative vector sum of the $\vec{\pT}$ of all selected and calibrated objects in the event, including a term to account for the momenta of soft particles that are not associated with any of the selected objects~\cite{PERF-2016-07}.
This soft term is calculated from ID tracks matched to the primary vertex, which makes it more resilient to contamination from \pileup interactions.

Scale factors are used to correct the efficiencies in simulation to match the efficiencies measured in data for the electron~\cite{EGAM-2018-01,EGAM-2021-01} and muon ~\cite{MUON-2018-03,MUON-2022-01} trigger, reconstruction, and identification and isolation criteria. Additional scale factors are also applied for the JVT requirement~\cite{PERF-2014-03}, for pileup and for the $b$-tagging efficiencies for jets that originate from the hadronization of $b$-quarks~\cite{FTAG-2018-01}, $c$-quarks~\cite{FTAG-2020-08}, and light quarks or gluons~\cite{FTAG-2019-02}.


%

\section{Event selection}
\label{sec:EventSelection}

The analysis targets \ttWjEW events in which the \Wboson boson decays leptonically, the same-charge top quark decays via a \Wboson boson to $\ell \nu b$ and the other top quark decays hadronically, giving rise to a dilepton signature including same-charge leptons, which can be exploited to suppress other dilepton backgrounds. Signal events are expected to present five jets: two \btagged jets originating from the two top quarks, two light jets originating from the hadronic top-quark decay and an additional spectator jet.

A preselection, defined to loosely match the expected signal topology, requires that the events contain exactly two same-charge leptons (electrons or muons). At least one lepton is required to be above $\pT(l_{0})> 28$~\GeV and to be trigger-matched, while $\pT(l_{1})> 20$~\GeV is required for the subleading lepton to reduce the fake-lepton background. The pseudorapidity of the muons must fulfill $|\eta|<2.5$, while $|\eta|<2.0$ is required for the electrons.
Electrons are required to satisfy the `tight' identification working point.
The event is required to contain at least three jets, with the leading jet having a transverse momentum $\pT(j_{0}) > 60$~\GeV. A requirement of the dilepton invariant mass, $m_{ll}$, to be greater than 30~\GeV is imposed to remove fake-lepton events. At least one of the jets must be $b$-tagged. These requirements are summarized in Table~\ref{tab:preSelSignalcuts}.

The signal is characterized by a topology with a forward jet and a high energy scale. The selection for the signal region requires at least four jets and the expected high energy scale is reflected in the requirements for a high $m_{ll} > 125$~\GeV and lepton $\pT(l_{0(1)}) > 80$~GeV (40~GeV) for the leading (subleading) lepton. The presence of a forward jet in the signal is exploited in the signal region definition, using the distribution of $\Delta\eta_{j}$: the difference in pseudorapidity between the most forward non-$b$-tagged jet and the jet with which it produces the highest invariant mass. These variables all help to provide separation between the \ttWjEW process and \ttW QCD\@.
In the case of di-electron events, events with invariant masses compatible with the \Zboson boson mass ($81.1~\mathrm{\GeV} < m_{ee} < 101.1~\mathrm{\GeV}$) are removed to reduce the contribution from electrons with misidentified charge. Events with fake leptons are reduced by a requirement on the scalar \pT sum of all selected jets ($\sum \pT (j) > 250$~\GeV). The fake leptons are primarily from the \ttbar process that has few jets from the harder scatter process and subsequently on average a lower $\sum \pT (j)$. Photon conversions are reduced by removing events with displaced vertices with a conversion radius (the radial distance from the collision point to conversion location) smaller than 50~mm associated with the electron track, if the conversion vertex is reconstructed. These requirements are also summarized in Table~\ref{tab:preSelSignalcuts}.

Dedicated control regions enriched in the dominant backgrounds are included in the fit to constrain these backgrounds from data. Table~\ref{tab:regionsDef} summarizes the definitions of the control regions used. There are seven control regions in total.
Two single-bin regions target the normalizations of backgrounds from photon conversions (Conv-CR) and misidentified electron charge (`charge-flips', CF-CR).
A region aiming to isolate \ttZ production (3L-CR (ttZ)) is defined by requiring at least one same-flavor opposite-charge (SFOC) lepton pair with $m_{ll}$ in a range compatible with the \Zboson boson mass and greater than four jets, of which at least two must be $b$-tagged. The region targeting diboson events (3L-CR) has the same selection except with lower multiplicity jet requirements of exactly three jets, of which exactly one must be $b$-tagged. To increase the separation between the diboson and $t\bar{t}Z$ processes, the two regions targeting them are binned in \HT (the scalar sum of transverse energy in the event excluding the \met). The other three regions (\crfourjlodeta, \crfourjhideta and \crthreej CRs) are binned according to the channel ($ee$, $e\mu$ and $\mu\mu$), the lepton charge, and the number of b-jets, for a total of 12 bins per region. The \crfourjlodeta region targets the normalization of the $t\bar{t}W$~QCD background at low $\Delta\eta$. The \crfourjhideta region targets the normalization of the $t\bar{t}W$~QCD and the fake-lepton background at high $\Delta\eta$ with the same $\Delta\eta$ requirement as in the signal region, but with orthogonal $H_\textrm{T}$, $\pT(l_{0})$, $\pT(l_{1})$ and $m_{ll}$ requirements to the signal region. The \crthreej CR constrains the overall non-$t\bar{t}W$~QCD backgrounds at high $m_{ll}$, in a phase-space close to the SR.\\

\begin{table}[htbp]
\centering
\caption{Summary of the definition of the preselection and signal regions. The $m_{ee}$ selection is used to exclude events with an electron-pair invariant mass compatible with that of the \Zboson boson mass ($81.1~\mathrm{\GeV} < m_{ee} < 101.1~\mathrm{\GeV}$).}
\begin{tabular}{lcc}
\toprule
& Preselection           & Signal Region \\
\midrule
Leptons                  & 2 same-charge leptons    & 2 same-charge leptons    \\
$\pT(l_{0})$             & $> 28$~\GeV             & $> 80$~\GeV             \\
$\pT(l_{1})$             & $> 20$~\GeV             & $> 40$~\GeV             \\
$|\eta(e)|$              & $< 2.0$                & $< 2.0$                \\
$|\eta(\mu)|$            & $< 2.5$                & $< 2.5$                \\
N (jets)                & $\geq 3$               & $\geq 4$               \\
N (b-jets)              & $\geq 1$               & $\geq 1$               \\
$\pT(j_{0})$             & ---                    & $> 60$~\GeV             \\
$m_{ll}$                 & $> 30$~\GeV             & $> 125$~\GeV            \\
$m_{ee}$                & ---                    & $\notin [81.1, 101.1]$~\GeV \\
$|\Delta \eta_{j}|$       & ---                    & $> 2.0$                \\
$\sum \pT (j)$          & ---                    & $> 250$~\GeV            \\
Conversion radius ($e$) & ---                    & $\geq 50$ mm           \\
\bottomrule
\end{tabular}
\label{tab:preSelSignalcuts}
\end{table}

\begin{landscape}
\begin{table}[htbp]
\centering
\scriptsize
\caption{Summary of the definition of the control regions. The `SR Veto' condition requires that an event does not satisfy the Signal Region requirements. The 3L-CR (\ttZ) $m_{ll}$ requirement applies to same-flavor opposite-charge (SFOC) leptons.}
\begin{tabular}{lccccccc}
\toprule
Control Regions             & \crthreej-CR & \crfourjlodeta-CR & \crfourjhideta-CR & Conv-CR                        & CF-CR              & 3L-CR (\ttZ)           & 3L-CR \\
\midrule
Leptons                    & 2 same-charge           & 2 same-charge         & 2 same-charge           & 2 same-charge ($ee$ or $e\mu$) & 2 same-charge $ee$ & 3 leptons             & 3 leptons \\
$\pt(l_{0})$           & $> 28$~\GeV            & $> 28$~\GeV          & $> 80$~\GeV            & $> 28$~\GeV                   & $> 50$~\GeV       & $> 28$~\GeV            & $> 28$~\GeV \\
$\pt(l_{1})$           & $> 20$~\GeV            & $> 20$~\GeV          & $> 20$~\GeV            & $> 20$~\GeV                   & $> 30$~\GeV       & $> 20$~\GeV            & $> 20$~\GeV \\
$\pt(l_{2})$           & ---                   & ---                 & ---                   & ---                          & ---              & $> 20$~\GeV            & $> 20$~\GeV \\
N(jets)                   & $= 3$                 & $\geq 4$            & $\geq 4$              & $\geq 4$                     & $\geq 3$         & $\geq 4$              & $= 3$ \\
N($b$-jets)               & $\geq 1$              & $\geq 1$            & $\geq 1$              & $\geq 1$                     & $\geq 1$         & $\geq 2$              & $= 1$ \\
$\pT(j_{0})$              & ---                   & $> 60$~\GeV          & $> 60$~\GeV            & $> 60$~\GeV                   & $> 60$~\GeV       & $> 60$~\GeV            & --- \\
$m_{ll}$            & $> 160$~\GeV           & $> 30$~\GeV          & $> 30$~\GeV            & $> 30$~\GeV                   & ---              & SFOC, $\in [81.1, 101.1]$~\GeV & --- \\
$m_{ee}$                  & $\notin [81.1, 101.1]$~\GeV & $\notin [81.1, 101.1]$~\GeV & $\notin [81.1, 101.1]$~\GeV & $\notin [81.1, 101.1]$~\GeV & $\in [81.1, 101.1]$~\GeV & --- & --- \\
$|\Delta\eta_{j}|$          & ---                   & $< 2.0$             & $> 2.0$         & ($< 2.0$ OR                     & ---              & ---                   & --- \\
$\sum \pT (j)$            & ---                   & ---                 & ---        &  $< 250$~\GeV)             & ---              & ---                   & --- \\
Conversion radius ($e$)   & ---                   & $\geq 50$ mm        & $\geq 50$ mm           & $< 50$ mm                   & $\geq 50$ mm     & ---                   & --- \\
\met                      & ---                   & ---                 & ---                   & ---                          & $< 100$~\GeV      & ---                   & --- \\
Veto condition            & ---                   & ---                 & SR Veto                & ---                          & ---    & ---                   & --- \\
\bottomrule
\end{tabular}
\label{tab:regionsDef}
\end{table}
\end{landscape}


%

\section{Background estimates}
\label{sec:NPFakes}

Backgrounds consisting of prompt leptons are estimated with the Monte Carlo templates described in Section~\ref{sec:DataMC}.
The normalizations of the $t \bar{t}Z$, $VV$ and $t \bar{t}W$ QCD processes are adjusted using unconstrained normalization factors that are determined by performing a likelihood fit to data across the CRs simultaneously with the SR.

Events containing CF electrons or fake leptons are significant backgrounds to the \ttWjEW process. These two categories are difficult to accurately model in simulations and are estimated from data.

The CF background is estimated by extracting charge-flip efficiency in di-electron events close to the $Z$ boson invariant mass peak. The selected events have at most one jet and no $b$-tagged jets. The corresponding efficiencies are binned in electron \pT and $\eta$ and then applied to opposite-charge events that have otherwise the same event selection applied as in the signal region. %

The fake-lepton event yield is estimated from data events using the Matrix Method (MM)~\cite{TOPQ-2010-01}, where contributions from `real leptons' (prompt leptons with correctly assigned charge), charge-flip, and photon conversion sources are subtracted using simulation.

The MM requires a sample of events with relaxed lepton criteria, which is achieved by removing the isolation requirements for both muons and electrons,
and using the loose electron identification selection.
The MM model input parameters are the real and fake electron and muon selection efficiencies for both the nominal and the relaxed criteria, which are extracted from a global simultaneous fit of the MM model.
The output of the MM model is a solution in terms of prompt and fake leptons, the latter of which are then used for the estimation of this background in the analysis.
The efficiencies are parameterized in lepton flavor, lepton charge, lepton transverse momentum, and number of $b$-tagged jets.
In the following, the fake background estimate is therefore provided grouped into four different templates corresponding to 1 or 2 b-jets and positive or negative lepton charges: Fakes 1b++/1b-{}-/2b++/2b-{}-.
The fake efficiency extraction regions are similar to the preselection but with inverted requirements on the invariant mass of the two leptons in three-jet events, and inverted maximum jet $\pT$ in four-jet events to ensure orthogonality to the other control and signal selections. A validation of the fake-lepton source composition is performed by confirming that agreement with data is observed in validation regions, defined by changing the three-jet selection to a four-jet selection.

The MM control regions have a relatively high predicted contamination of $t\bar{t}W$ events, ranging from 17\% in one-$b$-tagged events to 31\% in two-$b$-tagged events. To reduce the modeling dependence on this process, the MM parameters are determined as a function of the $t\bar{t}W$ cross section. The actual $t\bar{t}W$ cross section is later measured in situ and the fake-lepton background is automatically adjusted during the minimization of the likelihood used for the \ttWjEW measurement. This procedure properly accounts for the additional uncertainty induced by the $t\bar{t}W$-dependence of the fake-lepton estimate.

The background of leptons arising from photon conversions are estimated from simulated samples, isolating this contribution using generator-level information, and are normalized with a dedicated control region described in Section~\ref{sec:EventSelection}.


%

\section{Systematic uncertainties}
\label{sec:Systematics}

Systematic uncertainties from experimental sources and theoretical predictions can affect both the overall yield and the shape of the observables used in the fit for the statistical analysis.

The MC event yields are normalized to the data using the total integrated luminosity.
The uncertainty in the combined 2015--2018 integrated luminosity is 0.83\%~\cite{DAPR-2021-01}, obtained using the LUCID-2 detector~\cite{LUCID2} for the primary luminosity measurements, complemented by measurements using the ID and calorimeters.
The uncertainty in the reweighting of the MC pileup distribution to match the distribution in data is evaluated by varying the visible inelastic cross section.

For electrons and muons, the reconstruction, identification, isolation and trigger efficiency scale factors have associated uncertainties~\cite{EGAM-2021-02,MUON-2022-01}. The impact of the lepton energy scale and resolution uncertainty is evaluated by scaling or smearing the \pT of all leptons accordingly and re-applying the event selection~\cite{EGAM-2018-01,MUON-2018-03}.

Uncertainties are assigned to the scale and resolution of the soft track component of the \met. They are derived from differences between data and MC simulation in the measurement of the \pT balance between the hard and soft \met components~\cite{PERF-2016-07}.

The jet energy scale (JES) is calibrated using MC simulations, data taken during LHC runs, and test-beams as well as in situ information~\cite{ATL-PHYS-PUB-2015-036}. The uncertainty in the JES consists of many correlated components. To achieve a simplified setup with a minimum loss of correlation information, an eigenvector decomposition is used. A similar approach is used for the jet energy resolution (JER) uncertainty. The uncertainty associated with the JVT discriminant for pileup jet rejection is obtained by varying the efficiency correction factors~\cite{PERF-2014-03}.
Uncertainties are also considered in the $b$-tagging algorithm correction factors to the efficiencies measured from data which depend on the jet flavor and kinematic quantities of the jets. Additionally, flavor tagging extrapolation uncertainties for jets with $\pT$ above the calibration range are considered.

Alongside detector-related sources of systematic uncertainties, there are also uncertainties associated with the modeling of the MC predictions including their migrations between analysis regions. For prompt backgrounds that do not have free-floating normalizations in the fit, uncertainties in the predicted cross section are included as nuisance parameters: 20\% for $t\bar{t} H$~\cite{HIGG-2018-13}, 17\% for $tZq$~\cite{TOPQ-2018-02}, 20\% for other top quark processes and 30\% for `Other' minor processes. These values are chosen as constraints based on previous measurements of the relevant processes.

For the \ttWjEW process the uncertainty due to the ME calculation and the parton shower choices is estimated by comparing the nominal \SHERPA sample with the alternative \MGNLO + \PYTHIA[8] sample also at LO accuracy. The uncertainty due to the choice of parton shower, hadronization and underlying event model is estimated by comparing the difference between \POWHEG + \PYTHIA[8] and \POWHEG + \HERWIG[7] alternative samples and applying that difference to the nominal \SHERPA sample. To estimate the effect of missing higher order corrections, variations of $\mu_\mathrm{R}$ and $\mu_\mathrm{F}$ simultaneously up or down by a factor of two are applied. Uncertainties in the PDFs are evaluated by following the recommended PDF4LHC prescription~\cite{Butterworth:2015oua} and include variations of the \alphas scale and uncertainties related to the PDF choice. %
An uncertainty in the JES related to the fractions of quark- and gluon-initiated jets is included through variations relative to the nominal \ttWjEW sample, parameterized in jet $\eta$ and $p_T$.

For the \ttW QCD process, the uncertainty due to the generation of the hard-scattering process and the parton shower choices is estimated by comparing the nominal \SHERPA sample with an alternative \MGNLO + \PYTHIA[8] sample generated with up to one additional parton at NLO with FxFx merging~\cite{Frederix:2021agh}. The uncertainty due to the choice of parton shower, hadronization and underlying event model is estimated by computing the difference between the \POWHEG + \PYTHIA[8] and \POWHEG + \HERWIG[7] alternative samples and applying that difference to the nominal \SHERPA sample. The estimation of the effect of missing higher order corrections via variations of $\mu_\mathrm{R}$ and $\mu_\mathrm{F}$, the PDF uncertainties and the jet flavor uncertainties, are evaluated in the same way as in the \ttWjEW process.

For the $t\bar{t}Z$ and $tZq$ processes, the uncertainty due to the simulation of the initial-state radiation (ISR) is taken into account by varying the Var3c parameter of the \PYTHIA A14 tune. The uncertainty due to the parton shower choice is estimated by comparing the nominal \MGNLO + \PYTHIA[8.2] sample with an alternative \MGNLO sample interfaced with \Herwig. The PDF uncertainties and the jet flavor uncertainties are evaluated in the same way as in the \ttWjEW process.

As described in Section~\ref{sec:NPFakes}, several sources of uncertainties affecting the estimate of the fake rates are considered. These include the correlation of the lepton fakes with the $t\bar{t}W$ QCD process, the number of events in each \pT bin and the uncertainties in the background processes that enter into the likelihood fit to estimate the fake-lepton efficiency.


%

\section{Measurement under SM hypothesis and cross section upper limit extraction}
\label{sec:StatisticalAnalysis}

\begin{figure}[!htb]
\centering
\subfloat[]{
\includegraphics[width=0.32\textwidth]{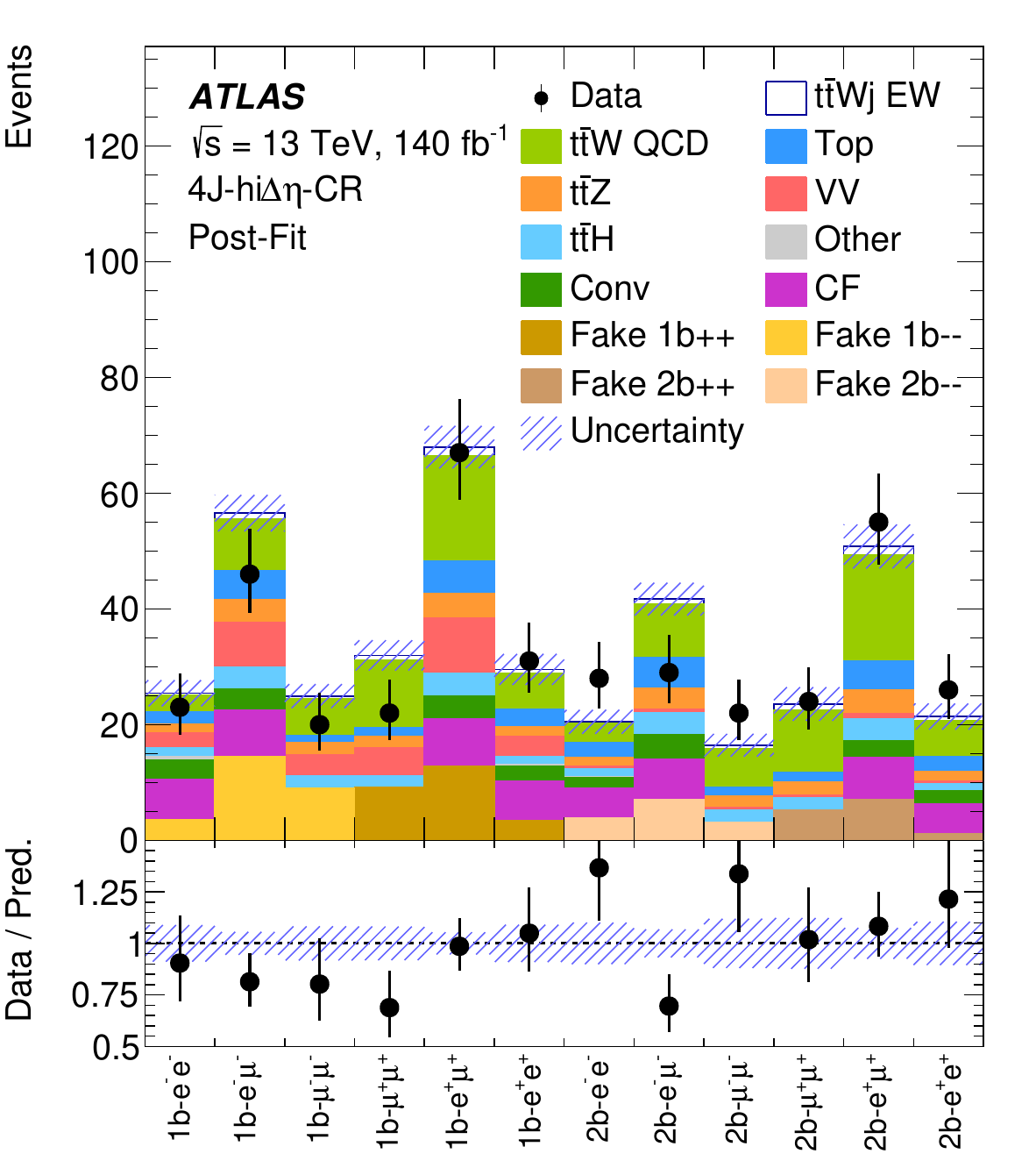}
}
\subfloat[]{
\includegraphics[width=0.32\textwidth]{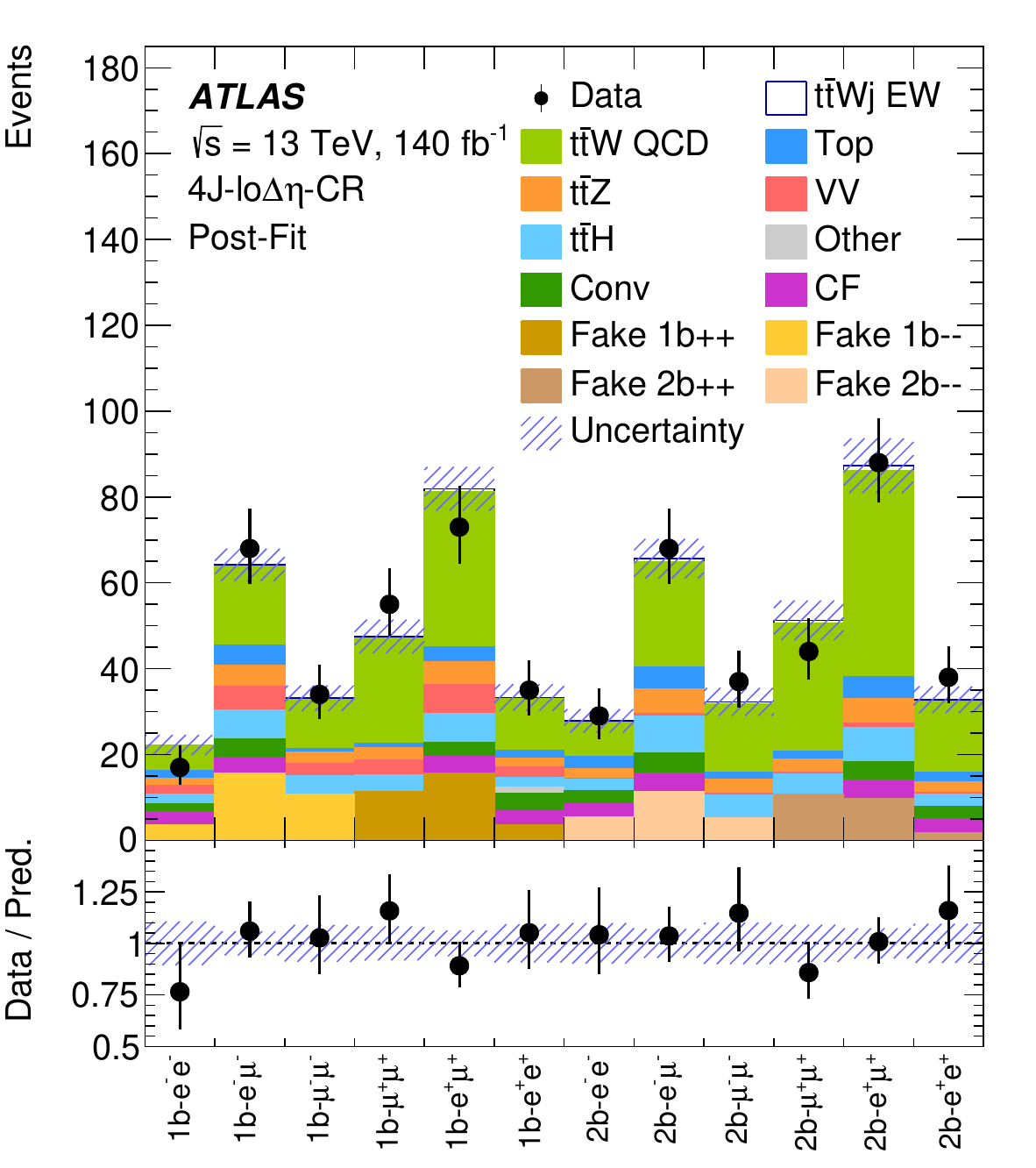}
}\\
\subfloat[]{
\includegraphics[width=0.32\textwidth]{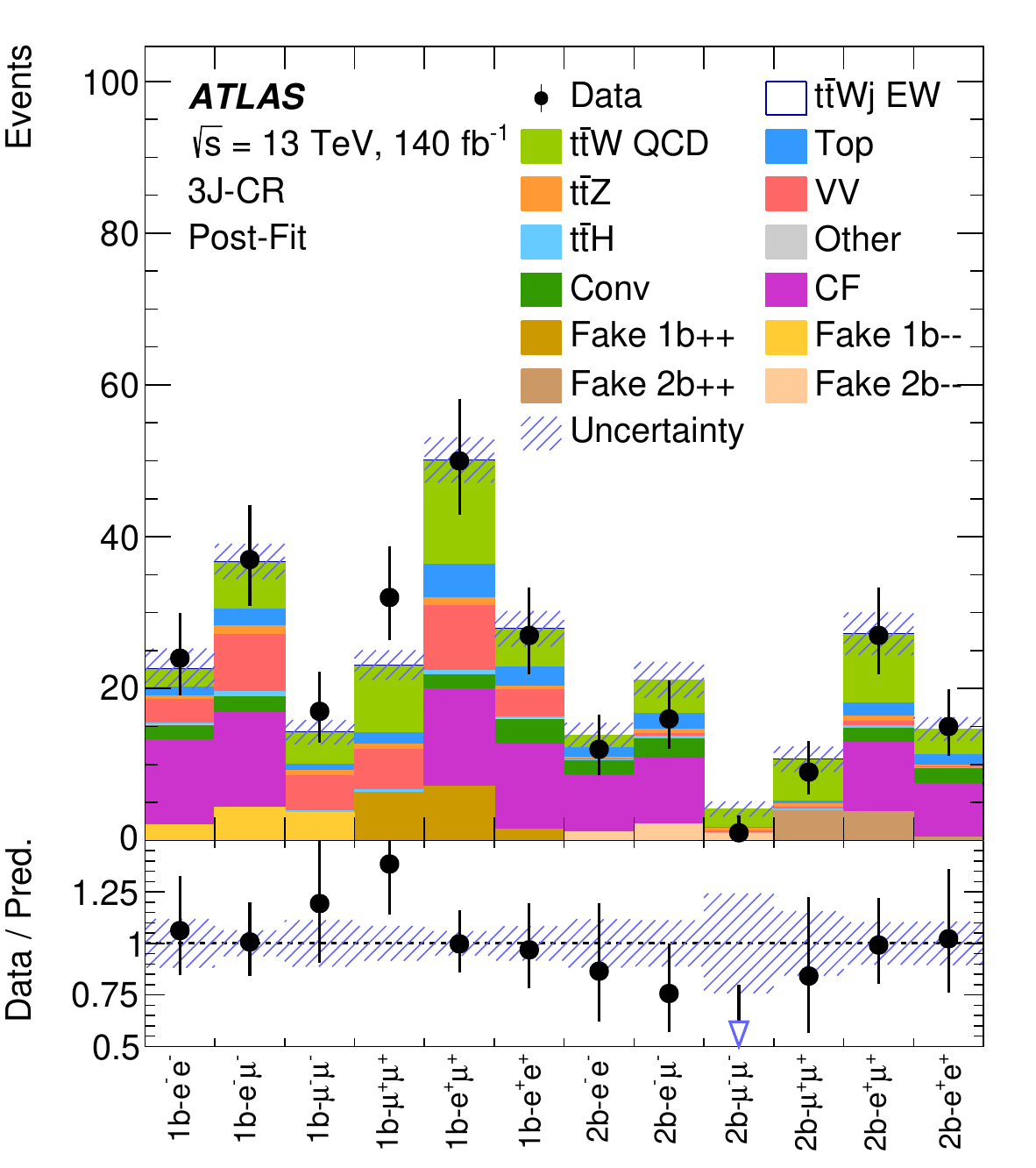}
}
\subfloat[]{
\includegraphics[width=0.32\textwidth]{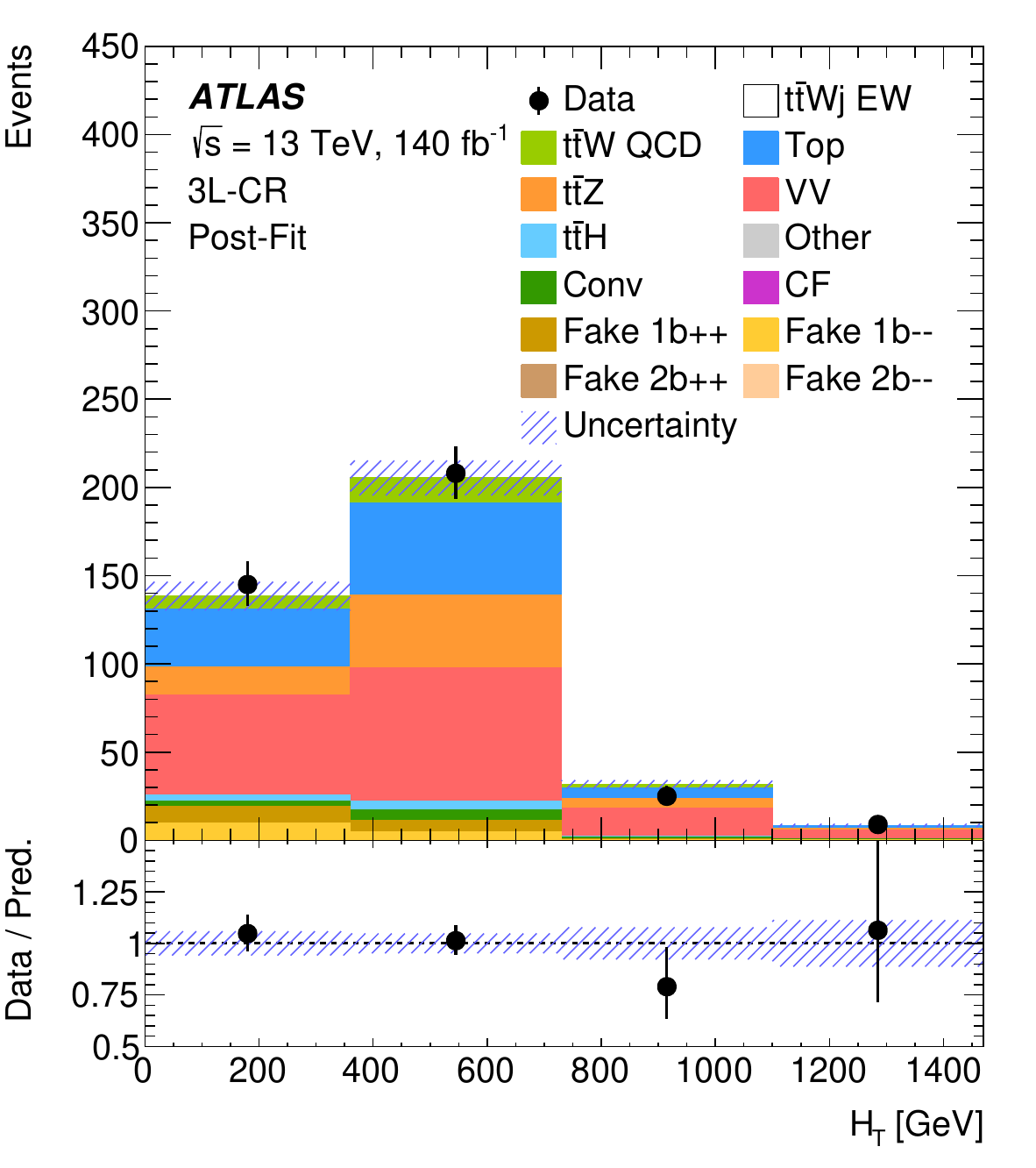}
}\\
\subfloat[]{
\includegraphics[width=0.32\textwidth]{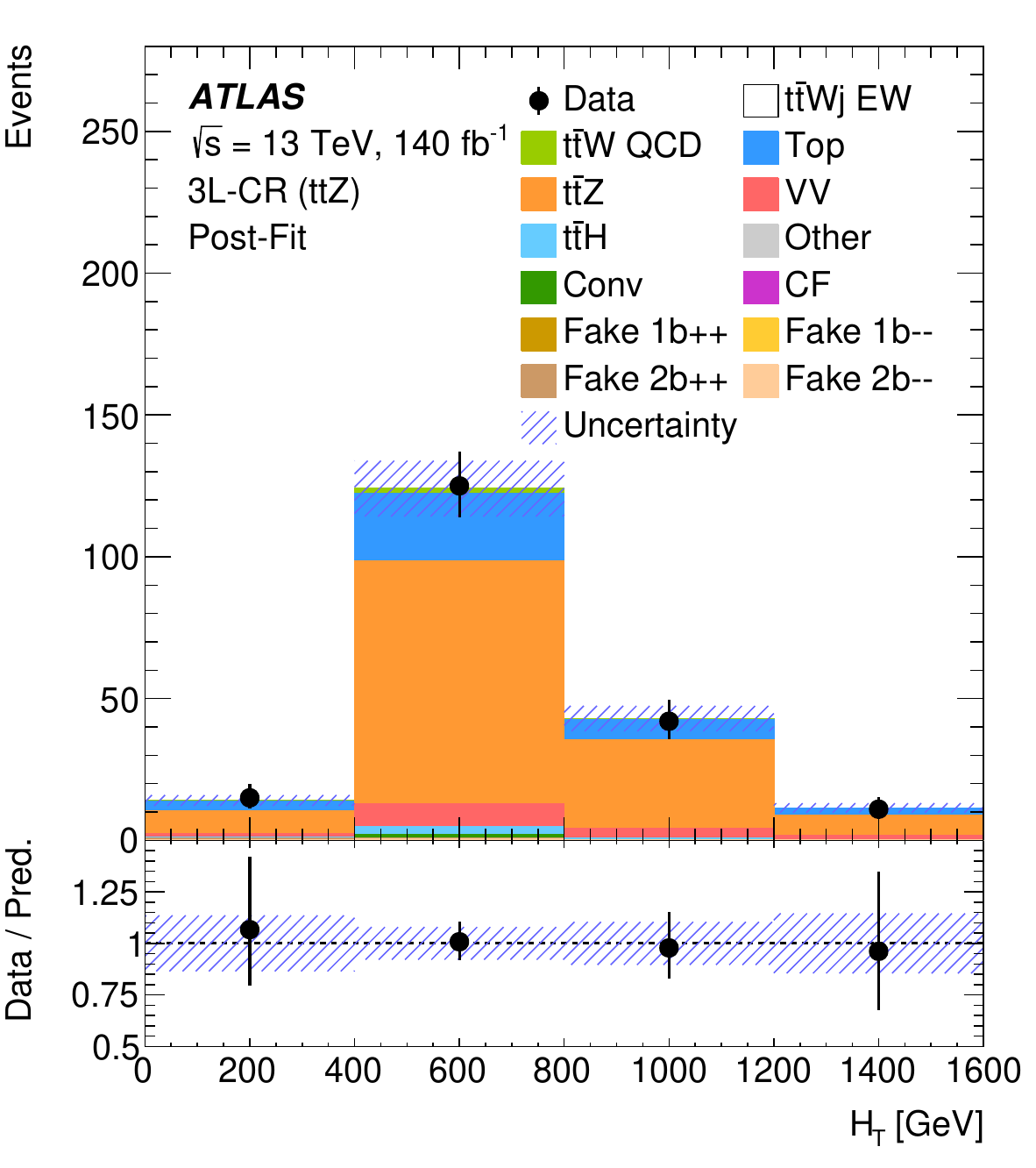}
}
\caption{Post-fit distributions in each of the analysis regions: (a) 4J-hi$\Delta\eta$-CR, (b) 4J-lo$\Delta\eta$-CR, (c) 3J-CR, (d) 3L-CR and (e) 3L-CR (ttZ). The shaded uncertainty bands include all statistical and systematic uncertainties. For the 3L-CR and 3L-CR (\ttZ) distributions, the last bin includes overflow events. For comparison with Table~\ref{tab:bgsigYields}, `Top' refers to `$tZq$' and `Other top backgrounds', `CF' refers to `charge-flip' events and `Conv' refers to `Conversions'.}
\label{fig:postfit_SM_CR1}
\end{figure}

\begin{figure}
\centering
\includegraphics[width=0.49\textwidth]{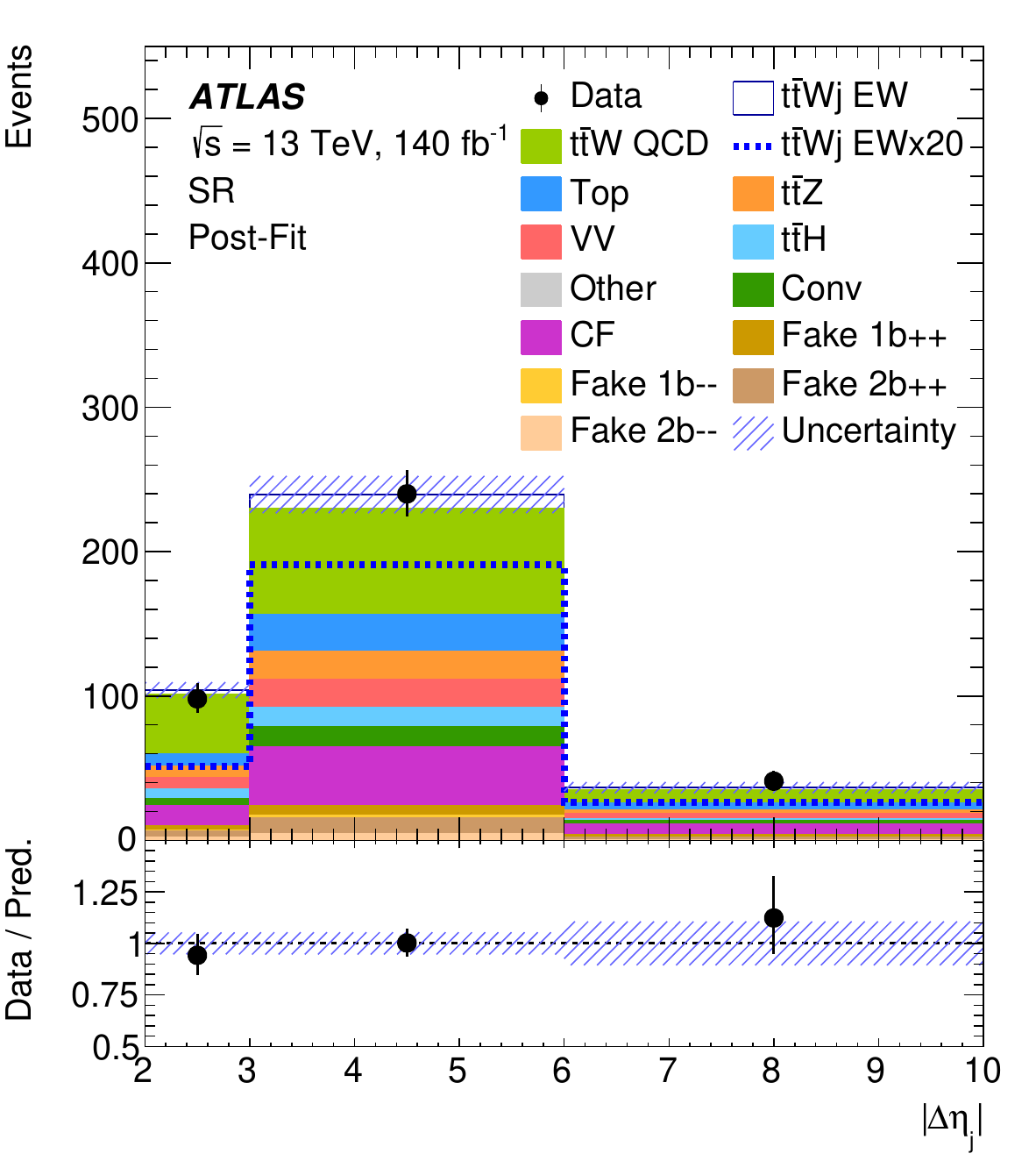}
\caption{Post-fit distribution of the $| \Delta\eta_j |$ observable in the signal region after the signal-plus-background fit to data. The shaded uncertainty band includes all statistical and systematic uncertainties. The binning starts from 2, following the signal region definition in Table \ref{tab:preSelSignalcuts}. For comparison with Table~\ref{tab:bgsigYields}, `Top' refers to `$tZ$' and `Other top backgrounds', `CF' refers to `charge-flip' events and `Conv' refers to `conversions'. In addition to the stacked histogram, the signal process is also shown as a dashed line overlaid, multiplied by a factor of 20.} \label{fig:postfit_SM_SR}
\end{figure}

The \ttWjEW production cross section, $\sigma_{\ttWjEW}$, is obtained using a binned profile-likelihood fit~\cite{Cowan:2010js}, using the \textsc{TRExFitter}~\cite{aly_2025_15524537} framework, to simultaneously fit the 46 CR bins and 3 SR bins, defined as a function of $| \Delta\eta_j |$. The expected signal and background distributions are modeled using template histograms from the MC simulated samples and the data-driven estimates as described in Section~\ref{sec:NPFakes}. The sum of signal and background contributions is fitted to the data. The likelihood function is constructed as a product of Poisson terms for the event yields and Gaussian functions to constrain nuisance parameters, which represent systematic uncertainties, across all bins included in the search. For uncertainties related to the finite number of simulated MC events, the Gaussian terms are replaced by Poisson terms, following the Beeston-Barlow `lite' method~\cite{Barlow:1993dm}.
The signal strength, which is the ratio of the observed \ttWjEW production cross section ($\sigma_{\textrm{obs}}$) to the theoretical \ttWjEW cross section ($\sigma_{\text{SM}}$) and denoted by $\mu_{\ttWjEW} = \sigma_{\textrm{obs}}/\sigma_{\text{SM}}$, is a free parameter in the profile-likelihood fit. The normalizations of the \ttZ, \ttW QCD, $VV$, photon conversion and electron charge misidentification processes are also free parameters in the profile-likelihood fit.
Table~\ref{tab:bgsigYields} shows the yields in the SR and CRs following a representative fit to data with the \ttWjEW signal normalization fixed to the SM expectation ($\mu_{\ttWjEW} = 1).$
Figures~\ref{fig:postfit_SM_CR1} and \ref{fig:postfit_SM_SR} show the post-fit distributions in the CRs and the SR, for a fit to data also including a free floating normalization factor for the \ttWjEW signal. The fitted normalization factors for the free floating processes in this fit are listed in Table~\ref{tab:norm}.

Within the experimental sensitivity, good agreement is found between the data and the SM prediction, resulting in an observed (expected) measurement of $\mu_{\ttWjEW} = 0.9^{+2.4}_{-3.0}$ $(1.0^{+2.4}_{-2.5})$. After applying the $\textrm{CL}_{s}$ method ~\cite{Read:2002hq} the observed (expected) $95\%$ $\textrm{CL}_{s}$ upper limit on the normalization factor is $\mu_{\ttWjEW}<5.3$ (4.8), which translates to an observed (expected) upper limit on the cross section of $\sigma_{\ttWjEW}<251$ fb (230~fb), to be compared with the expected SM cross section of $47.7^{+1.9}_{-1.4}$ fb~\cite{Frederix:2021agh}.
The measurement is overall dominated by systematic uncertainties, detailed in Table~\ref{tab:impact_muttWjEW}. The leading sources of uncertainty in the fitted value of $\mu_{\ttWjEW}$ are the normalization and modeling of the $t\bar{t}W$ QCD process, uncertainties related to the jet modeling, and the estimate of the fake-lepton contributions.

\sisetup{
separate-uncertainty = true,
table-align-uncertainty = true,
detect-weight = true,
detect-inline-weight = math
}

\begin{table}[htbp]
\centering
\caption{Observed and expected event yields in the signal and control regions, obtained for an integrated luminosity of 140~\ifb after performing fit to the data with the \ttWjEW normalization fixed to the SM expectation ($\mu_{\ttWjEW} = 1$) with corresponding uncertainties. The indicated errors include the Monte Carlo statistical uncertainties and all other systematic uncertainties. The quadrature sum of uncertainties is not equal to the total uncertainty because of correlations introduced by the fit. A dash (-----) indicates event yields smaller than 0.1 events.}
\resizebox{\textwidth}{!}{
\begin{tabular}{
l
S@{\,±\,}S
S@{\,±\,}S
S@{\,±\,}S
S@{\,±\,}S
S@{\,±\,}S
S@{\,±\,}S
S@{\,±\,}S
S@{\,±\,}S
}
\toprule
& \multicolumn{2}{c}{SR}
& \multicolumn{2}{c}{4J-lo$\Delta\eta$-CR}
& \multicolumn{2}{c}{4J-hi$\Delta\eta$-CR}
& \multicolumn{2}{c}{3J-CR}
& \multicolumn{2}{c}{3L-CR (ttZ)}
& \multicolumn{2}{c}{3L-CR}
& \multicolumn{2}{c}{Conv-CR}
& \multicolumn{2}{c}{CF-CR} \\
\midrule
${t\bar{t}W}_{QCD}$ & 121 & 26 & 250 & 40 & 106 & 22 & 67 & 13 & 2.6 & 0.6 & 24 & 5 & 0.93 & 0.25 & 7.6 & 1.4 \\
$tZq$ & 7.6 & 1.4 & 2.20 & 0.35 & 9.2 & 2.2 & 7.5 & 1.2 & 23 & 4 & 68 & 11 & \multicolumn{2}{c}{\hspace{0.1em}\text{-----}} & 0.88 & 0.22 \\
\ttZ & 31 & 4 & 43 & 8 & 30 & 4 & 7.2 & 1.7 & 133 & 15 & 66 & 14 & 0.26 & 0.15 & 2.2 & 0.4 \\
$VV$ & 30 & 6 & 26 & 5 & 34 & 7 & 34 & 6 & 14.1 & 2.7 & 152 & 27 & 0.23 & 0.04 & 3.8 & 0.7 \\
\ttH & 21 & 4 & 59 & 12 & 29 & 6 & 3.9 & 0.9 & 3.7 & 0.8 & 9.3 & 1.9 & 0.27 & 0.06 & 2.3 & 0.5 \\
Other top bkg. & 30 & 6 & 30 & 6 & 27 & 6 & 12.8 & 3.0 & 13.2 & 2.7 & 25 & 5 & 14.1 & 3.2 & 2.4 & 0.6 \\
Other   & 0.25 & 0.07 & 1.6 & 0.5 & 1.04 & 0.30 & 0.20 & 0.06 & \multicolumn{2}{c}{\hspace{0.9em}\text{-----}} & \multicolumn{2}{c}{\hspace{0.9em}\text{-----}} & 0.34 & 0.10 & 6.6 & 1.9 \\
Conversions & 20 & 7 & 28 & 10 & 24 & 8 & 17 & 6 & 1.5 & 0.5 & 9 & 4 & 27 & 8 & 13 & 5 \\
Charge-flips & 62 & 9 & 29 & 4 & 55 & 8 & 80 & 12 & \multicolumn{2}{c}{\hspace{0.9em}\text{-----}} & \multicolumn{2}{c}{\hspace{0.9em}\text{-----}} & 0.68 & 0.11 & 68 & 10 \\
Fakes 1b$++$   & 11 & 5 & 30 & 10 & 25 & 8 & 15 & 7 & \multicolumn{2}{c}{\hspace{0.9em}\text{-----}} & 16 & 5 & 0.8 & 0.6 & 1.6 & 1.4 \\
Fakes 1b$--$   & 3.5 & 1.6 & 30 & 7 & 27 & 5 & 10.0 & 3.1 & \multicolumn{2}{c}{\hspace{0.9em}\text{-----}} & 15.0 & 3.4 & 0.9 & 0.4 & 3.2 & 1.4 \\
Fakes 2b$++$   & 16 & 6 & 22 & 10 & 13 & 6 & 8.0 & 3.1 & 0.9 & 0.7 & \multicolumn{2}{c}{\hspace{0.9em}\text{-----}} & 0.71 & 0.31 & 0.65 & 0.30 \\
Fakes 2b$--$   & 7.1 & 2.9 & 22 & 8 & 14 & 5 & 4.1 & 2.0 & 0.66 & 0.10 & \multicolumn{2}{c}{\hspace{0.9em}\text{-----}} & 1.9 & 0.5 & 1.8 & 0.6 \\
\midrule
\ttWjEW ($\mu_{\ttWjEW} = 1$) & 15.7 & 2.6 & 7.3 & 2.6 & 11.5 & 1.9 & 1.31 & 0.26 & \multicolumn{2}{c}{\hspace{0.9em}\text{-----}} & 0.124 & 0.028 & \multicolumn{2}{c}{\hspace{0.9em}\text{-----}} & 0.53 & 0.13 \\
\midrule
Total & 376 & 20 & 584 & 23 & 405 & 16 & 268 & 12 & 192 & 14 & 384 & 19 & 48 & 7 & 114 & 8 \\
\midrule
Data & \multicolumn{2}{c}{\llap{379}\phantom{0000000}} & \multicolumn{2}{c}{\llap{586}\phantom{0000000}} & \multicolumn{2}{c}{\llap{393}\phantom{0000000}} & \multicolumn{2}{c}{\llap{267}\phantom{0000000}}
& \multicolumn{2}{c}{\llap{193}\phantom{0000000}} & \multicolumn{2}{c}{\llap{387}\phantom{0000000}} & \multicolumn{2}{c}{\llap{49}\phantom{0000000}} & \multicolumn{2}{c}{\llap{119}\phantom{0000000}} \\
\bottomrule
\end{tabular}
}
\label{tab:bgsigYields}
\end{table}

\begin{table}[htbp]
\centering
\caption{Signal and background normalization factors measured in the profile-likelihood signal-plus-background fit to the data. All sources of statistical and systematic uncertainties are included.}
{
\renewcommand{\arraystretch}{1.4}
\begin{tabular}{l c}
\toprule
Process & Normalization factor ($\mu$) \\
\midrule
\ttWjEW        & $0.9\;^{+2.4}_{-3.0}$ \\
\ttZ           & $0.94\;^{+0.14}_{-0.13}$ \\
${t\bar{t}W}_{QCD}$           & $1.06\;^{+0.28}_{-0.26}$ \\
$VV$           & $1.11\;^{+0.20}_{-0.20}$ \\
Conversions    & $1.12\;^{+0.37}_{-0.34}$ \\
Charge-flips   & $0.92\;^{+0.14}_{-0.14}$ \\
\bottomrule
\end{tabular}
}
\label{tab:norm}
\end{table}

\sisetup{
explicit-sign,
table-align-uncertainty = true,
detect-weight = true
}

\begin{table}[htbp]
\centering
\caption{Grouped impact of systematic uncertainties in the \ttWjEW signal strength, $\mu_{\ttWjEW}$, after the profile-likelihood signal-plus-background fit to the data. The impact of each uncertainty is assessed by fixing the relevant fit parameters for a given uncertainty and re-fitting to data. The reduction in the total uncertainty in this modified fit gives the quoted impact. The quadrature sum of individual groups of uncertainties is not equal to the total uncertainty because of correlations introduced by the fit.}
\renewcommand{\arraystretch}{1.3}
\begin{tabular}{l
S[table-format=+0.4]
@{\,/\,}
S[table-format=+1.7]}
\toprule
Uncertainty Source & \multicolumn{2}{c}{$\Delta\mu_{\ttWjEW}$} \\
\midrule
Total Statistical and Systematic          & {+}2.4     & {--}3.0 \\
Total Systematic                          & {+}2.2  & {--}2.8 \\
Total Statistical                         & {+}0.9  & {--}0.9 \\
\midrule
${t\bar{t}W}_{\textrm{QCD}}$ normalization         & {+}1.6  & {--}2.2 \\
Jets and \met                              & {+}1.6  & {--}2.2 \\
\ttWjEW modeling                         & {+}0.4 & {--}1.5 \\
${t\bar{t}W}_{\textrm{QCD}}$ modeling                            & {+}0.6 & {--}1.2 \\
MC statistical                            & {+}0.7 & {--}0.9 \\
Fake leptons                              & {+}0.5 & {--}0.7 \\
Pileup                                   & {+}0.2 & {--}0.5 \\
\ttZ normalization                        & {+}0.3 & {--}0.4 \\
\ttZ modeling                            & {+}0.23 & {--}0.27 \\
Background cross sections                 & {+}0.18 & {--}0.31 \\
Electron ID and isolation                 & {+}0.09 & {--}0.09 \\
$b$-tagging                               & {+}0.09 & {--}0.09\\
Muons                                     & {+}0.07 & {--}0.07 \\
$tZq$ modeling                            & {+}0.07 & {--}0.07 \\
Luminosity                                & {+}0.034  & {--}0.034 \\
Electron energy scale and resolution      & {+}0.024 & {--}0.024 \\
JVT                                       & \multicolumn{2}{c}{$< 0.01$} \\
\bottomrule
\end{tabular}
\label{tab:impact_muttWjEW}
\end{table}

\FloatBarrier


%

\section{Effective field theory interpretation}
\label{sec:EFTInterpretation}

New physics could affect the \ttWjEW process at high energy scales and can be parameterized in a model-independent way in the context of an EFT\@. The \ttWjEW cross section is increased by non-zero Wilson coefficients for operators such as \Opt and \OHQone~\cite{Dror:2015nkp}. These operators also contribute to modifications of the \ttZ process~\cite{TOPQ-2020-20}. Therefore, along with an assessment of the effect of \Opt and \OHQone on \ttWjEW production, the complementarity with \ttZ is exploited in a fit assessing contributions from the two EFT operators to both physics processes.

\begin{figure}[h]
\centering
\includegraphics[width=0.49\textwidth]{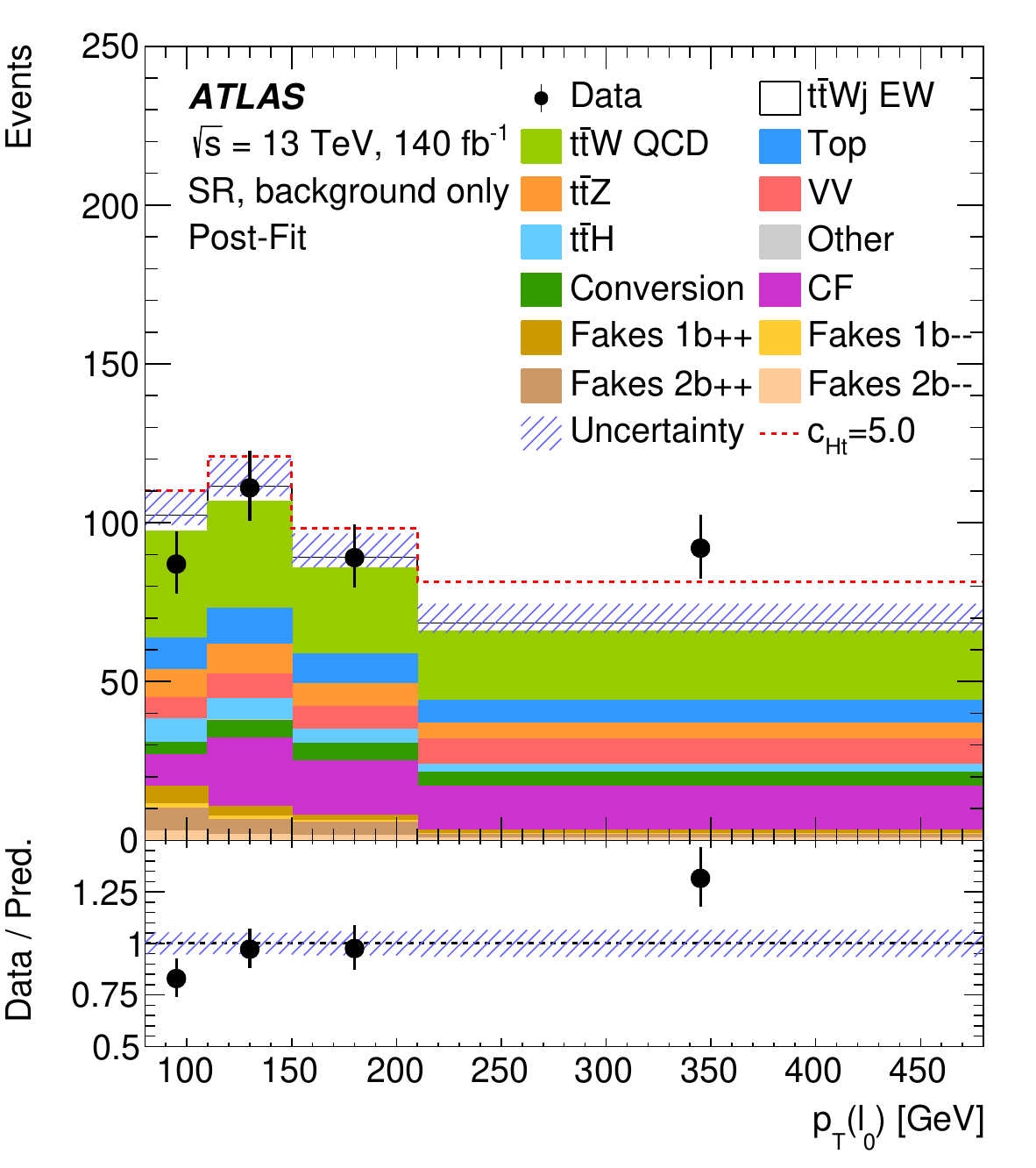}
\caption{Distribution of the EFT-sensitive observable $p_{\textrm{T}}(l_0)$ in the signal region, after a fit to data of the SM processes in the control regions only. The dashed line represents the expected total distribution of an EFT signal stacked on top of the SM processes, with $\cpt=5$ for illustration purposes. The shaded uncertainty band includes all statistical and systematic uncertainties. The highest bin is inclusive and includes all events above the upper bin edge.
}
\label{fig:SR_SM_fitEFTBinning}
\end{figure}

As the effects of \Opt and \OHQone grow with the energy scale of the scattering process, the signal region is re-binned in four bins of $\pT(l_0)$ to maximize the sensitivity to the EFT operators. Figure~\ref{fig:SR_SM_fitEFTBinning} shows the $\pT(l_0)$ distribution in the signal region, comparing observed events to the SM prediction normalized in a fit to the control regions only. The subsequent EFT fits are otherwise performed identically to the SM measurement, with the exception that when an EFT operator is considered to affect the \ttWjEW (\ttZ) process, the floating normalization factor for that process is replaced with a dedicated cross section uncertainty nuisance parameter of size $^{+4\%}_{-3\%}$~\cite{Frederix:2017wme,Frederix:2021agh} ($\pm 9.5\%$~\cite{Kulesza:2020nfh}). Additionally, a new free-floating parameter is introduced to parameterize the effect of the EFT operator on the \ttWjEW (\ttZ) normalization in each bin of the analysis. The assumed energy scale is set to  $\Lambda = 1$~\TeV.
Several fit configurations are considered for the EFT interpretations. One configuration allows EFT variations only on \ttWjEW production and not \ttZ, and another configuration allows EFT variations on both \ttWjEW and \ttZ. For both these configurations, interpretations are made considering both each EFT operator alone, and both operators simultaneously. This combination of interpretations is designed to disentangle the sensitivity from each process to each operator and probes the nature of their complementarity. For these configurations the EFT effects are applied to the relevant processes in the rebinned SR and all CRs.

The asymptotic likelihood approach relies on Wilks' theorem~\cite{Wilks:1938dza} for the extraction of the confidence intervals using the properties of a $\chi^{2}$ distribution~\cite{Cowan:2010js}. However, in the EFT fits the \ttWjEW process has non-negligible quadratic dependencies on Wilson coefficients, which lead to violations of Wilks' theorem~\cite{Bernlochner:2022oiw} and over- or under-coverage of the true intervals. An alternative method for extracting confidence intervals is employed in which pseudo-datasets are generated at different values of a Wilson coefficient and fitted individually. From these fits, test-statistic distributions are constructed to derive the final confidence intervals.

\begin{figure}[b!]
\centering
\subfloat[]{\includegraphics[width=0.49\textwidth]{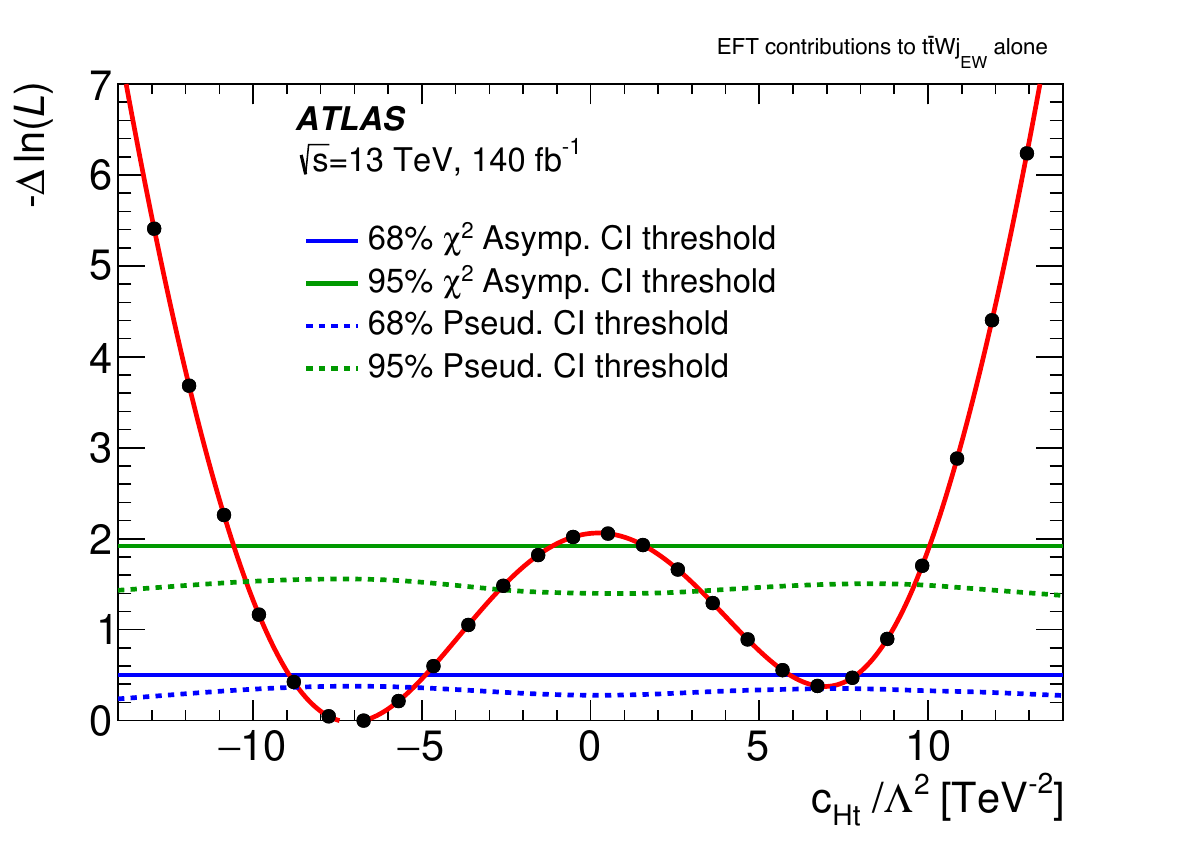}\label{fig:EFT_ttWjEW_1Ds_a}}
\subfloat[]{\includegraphics[width=0.49\textwidth]{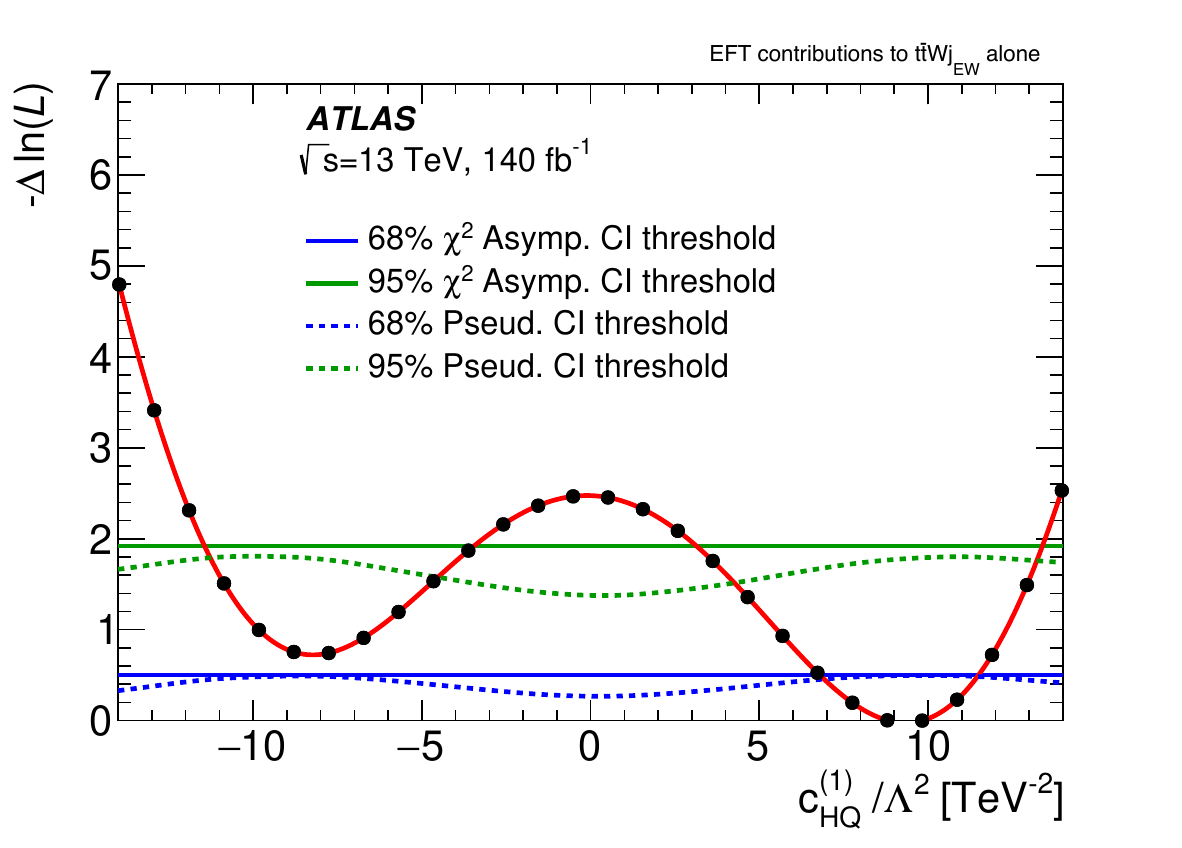}\label{fig:EFT_ttWjEW_1Ds_b}}
\caption[dummy]{Difference in the log likelihood as a function of \subref{fig:EFT_ttWjEW_1Ds_a} \cpt and \subref{fig:EFT_ttWjEW_1Ds_b} \cHQone Wilson coefficients, as determined in the EFT fit to data, considering the impact of each coefficient independently on the \ttWjEW process. The solid lines show the asymptotic $\chi^{2}$ 68\% and 95\% thresholds (Asymp.) for calculating confidence intervals. The dashed lines show alternative estimations of the 68\% and 95\% confidence interval thresholds using pseudo-datasets (Pseud.), as described in the text.} \label{fig:EFT_ttWjEW_1Ds}
\end{figure}

\begin{figure}
\centering
\subfloat[]{\includegraphics[width=0.49\textwidth]{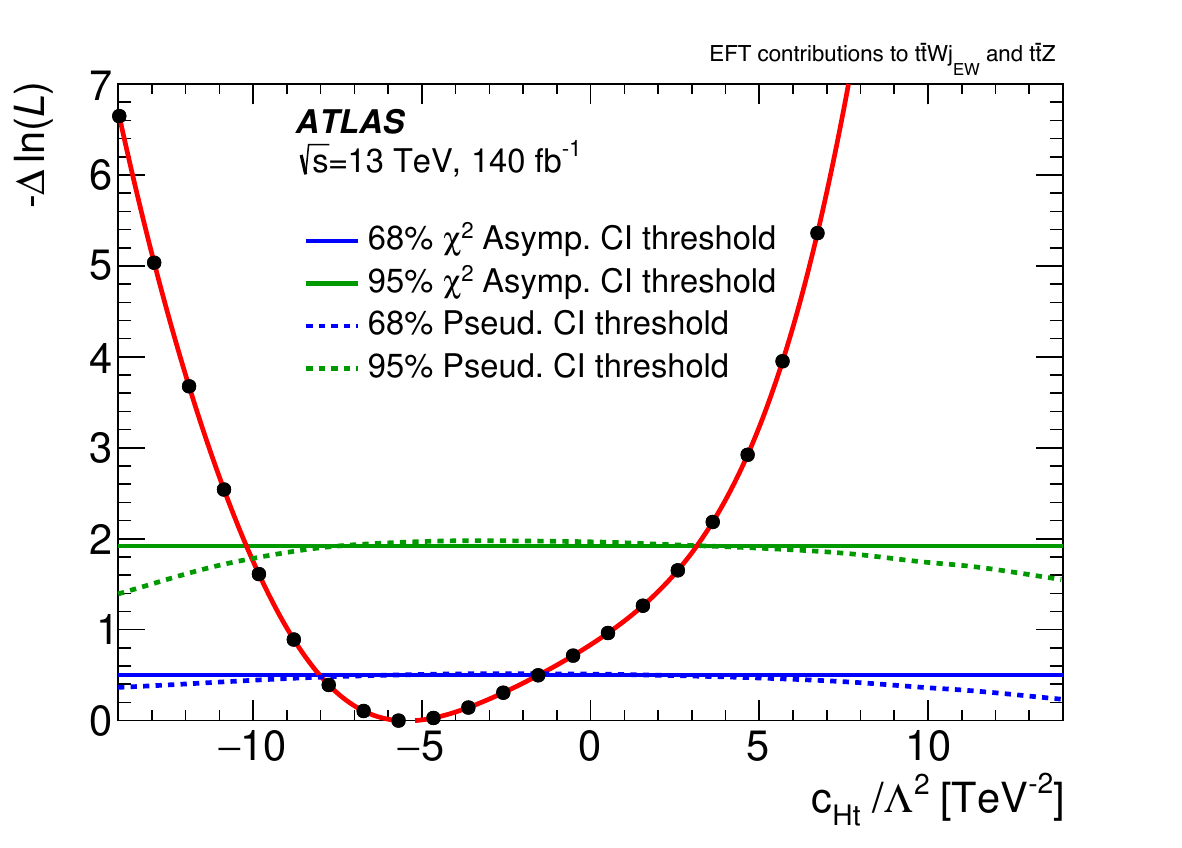}\label{fig:EFT_ttWjEW_ttZ_1Ds_a}}
\subfloat[]{\includegraphics[width=0.49\textwidth]{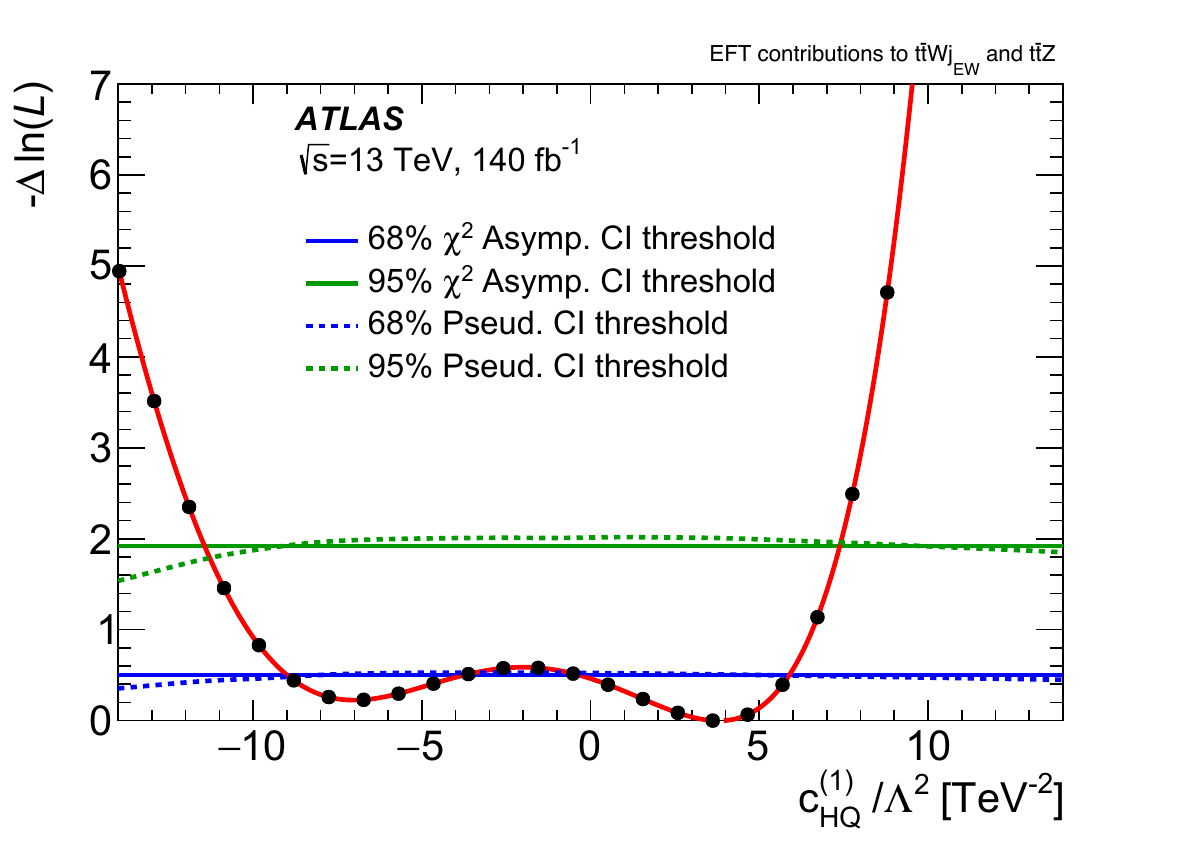}\label{fig:EFT_ttWjEW_ttZ_1Ds_b}}
\caption[dummy]{Difference in the log likelihood as a function of \subref{fig:EFT_ttWjEW_ttZ_1Ds_a} \cpt and \subref{fig:EFT_ttWjEW_ttZ_1Ds_b} \cHQone Wilson coefficients, as determined in the EFT fit to data, considering the corresponding EFT contributions from one coefficient at a time but simultaneously on both the the \ttWjEW and \ttZ processes. The solid lines show the asymptotic $\chi^{2}$ 68\% and 95\% thresholds (Asymp.) for calculating confidence intervals. The dashed lines show alternative estimations of the 68\% and 95\% confidence interval thresholds using pseudo-datasets (Pseud.), as described in the text.} \label{fig:EFT_ttWjEW_ttZ_1Ds}
\end{figure}

Figure~\ref{fig:EFT_ttWjEW_1Ds} shows likelihood scans considering the impact of \Opt and \OHQone independently on the \ttWjEW process alone. The one-dimensional $\chi^{2}$ threshold at $-\Delta\ln(L)=0.5$, which relies on Wilks' theorem, results in compatible but looser confidence intervals than determined from the alternative threshold using pseudo-datasets. Figure~\ref{fig:EFT_ttWjEW_ttZ_1Ds} shows likelihood scans considering the impact of \Opt and \OHQone independently on both the \ttWjEW and \ttZ processes.  The larger linear contributions of the EFT operators to the \ttZ process mitigate coverage changes due to the violation of Wilks' theorem.

\begin{figure}
\centering
\subfloat[]{\includegraphics[width=0.49\textwidth]{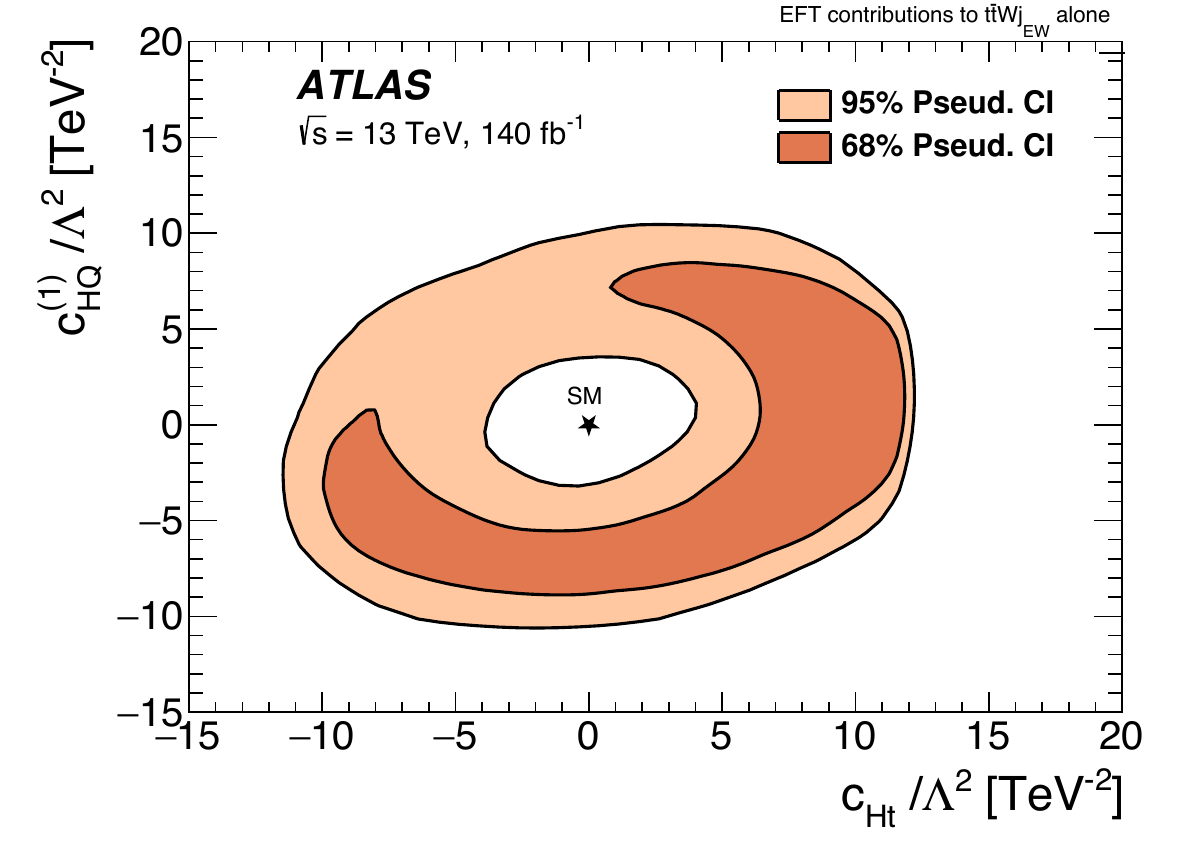}\label{fig:EFT_ttWjEW_ttWjEW_ttZ_2D_a}}
\subfloat[]{\includegraphics[width=0.49\textwidth]{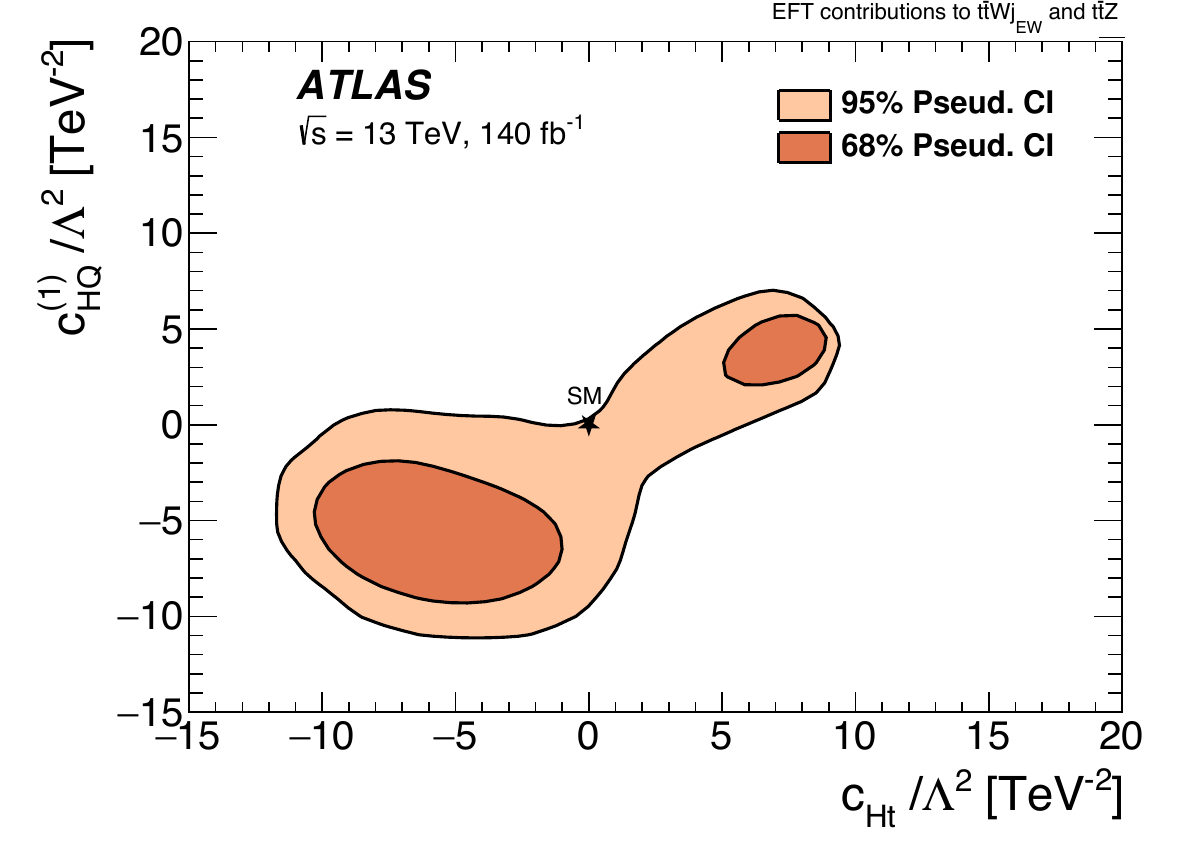}\label{fig:EFT_ttWjEW_ttWjEW_ttZ_2D_b}}
\caption[dummy]{Two-dimensional 68\% and 95\% confidence intervals for the simultaneous EFT fit to data of the Wilson coefficients \cpt and \cHQone, considering their impact on \subref{fig:EFT_ttWjEW_ttWjEW_ttZ_2D_a} the \ttWjEW process alone or \subref{fig:EFT_ttWjEW_ttWjEW_ttZ_2D_b} on both the \ttWjEW and \ttZ processes. The confidence intervals are determined using pseudo-datasets (Pseud.).} \label{fig:EFT_ttWjEW_ttWjEW_ttZ_2D}
\end{figure}

\begin{figure}
\centering
\includegraphics[width=\textwidth]{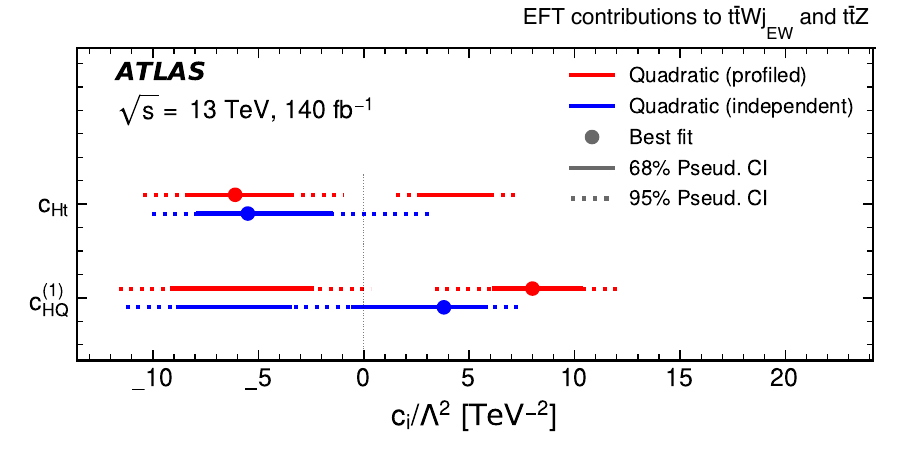}
\caption{95\% confidence intervals determined using pseudo-datasets (Pseud.) for the \cpt and \cHQone Wilson coefficients from fits to both \ttWjEW and \ttZ. The intervals are obtained from both fits to each EFT operator independently (independent), and from two-dimensional fits to both operators simultaneously (profiled).} \label{eft_summary}
\end{figure}

\begin{figure}
\centering
\includegraphics[width=\textwidth]{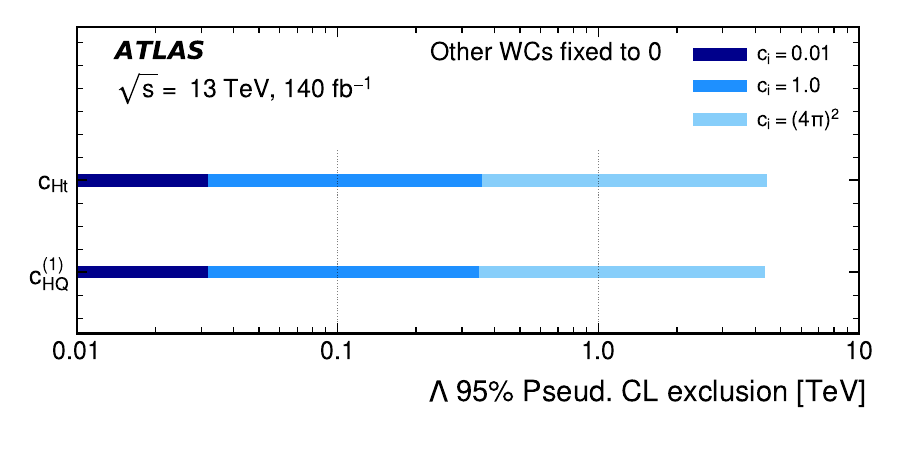}
\caption{95\% confidence level exclusion limits on the new physics scale $\Lambda$ for different assumptions on the value of the (dimensionless) Wilson Coefficients \cpt and \cHQone. The limits are determined for each EFT operator independently by converting the average magnitude of the up and down uncertainties from the pseudo-datasets (Pseud.) fits to both \ttWjEW and \ttZ.}
\label{eft_summary_lambda}
\end{figure}

Figure~\ref{fig:EFT_ttWjEW_ttWjEW_ttZ_2D} shows the full two-dimensional 95\% confidence intervals determined using pseudo-datasets when fitting both \cpt and \cHQone simultaneously.
The numerical results of all fits are summarized in Figures~\ref{eft_summary} and~\ref{eft_summary_lambda} and are detailed in Table~\ref{tab:all_EFT_limits_95}. The EFT fits use both linear and quadratic terms and consider either both EFT operators simultaneously, \cpt + \cHQone (profiled), or only individual EFT Wilson coefficients (independent). The fits consistently prefer a non-zero value for the EFT Wilson coefficients, due to the difference between data and MC observed in the highest \pt bin of Figure~\ref{fig:SR_SM_fitEFTBinning}, which can be accommodated by the impacts of \Opt and \OHQone. In the most complex fit, considering the impact of both \Opt or \OHQone on the \ttWjEW and \ttZ processes, the $95\%$ confidence interval contains the SM expectation.

\begin{figure}[b!]
\centering
\subfloat[]{\includegraphics[width=0.49\textwidth]{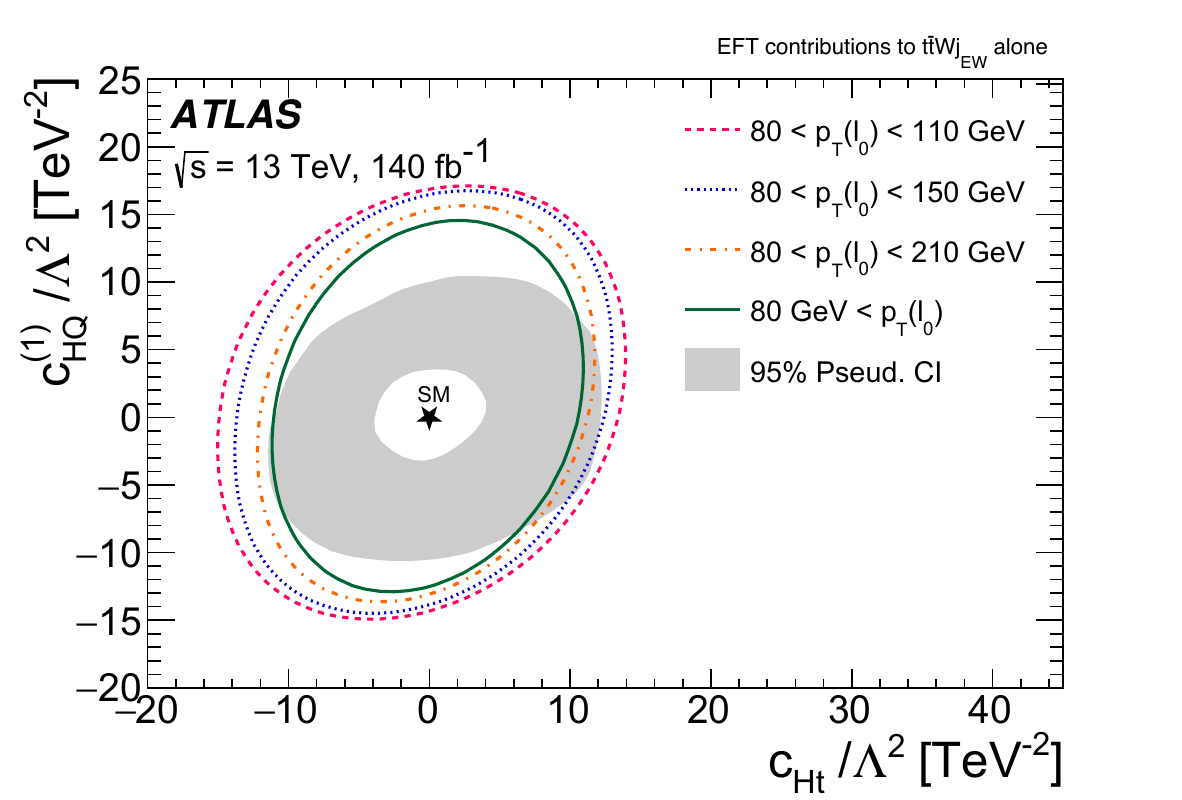}\label{fig:energygrowth_ttWjttZ_a}}
\subfloat[]{\includegraphics[width=0.49\textwidth]{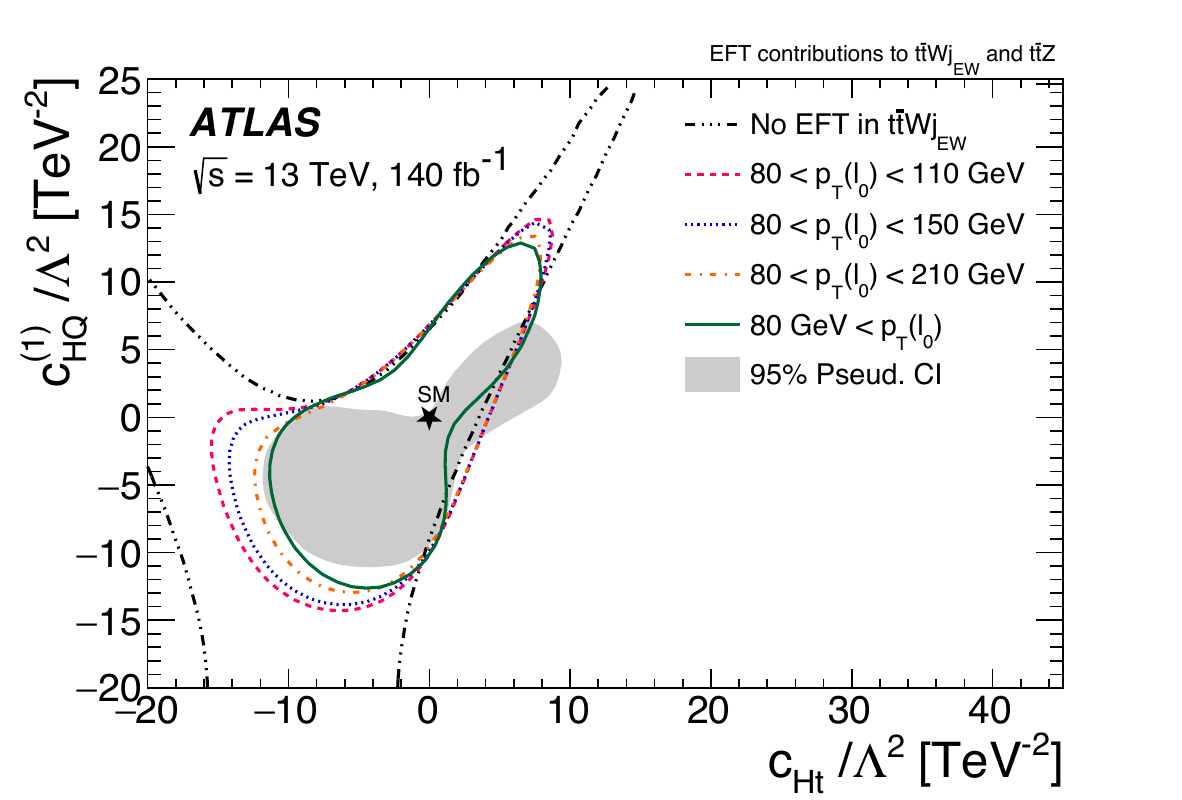}\label{fig:energygrowth_ttWjttZ_b}}
\caption[dummy]{Two-dimensional 68\% and 95\% confidence intervals for the simultaneous EFT fit to data of the Wilson coefficients \cpt and \cHQone, considering their impact on \subref{fig:energygrowth_ttWjttZ_a} the \ttWjEW process alone or \subref{fig:energygrowth_ttWjttZ_b} on both the \ttWjEW and \ttZ processes. The dashed, dotted and solid line contours use the asymptotic likelihood approach, and highlight the energy-growth behavior of the EFT operators. On \subref{fig:energygrowth_ttWjttZ_b}, the outermost open contour shows the confidence interval when considering the EFT operators on \ttZ only. Subsequently, on both \subref{fig:energygrowth_ttWjttZ_a} and \subref{fig:energygrowth_ttWjttZ_b} the closed contours consider the EFT operators also on \ttWjEW in the four \pT bins in the signal region. The outer closed contour uses only the first \pT bin and moving inwards the contours include higher energy information for the 2-bin, 3-bin and 4-bins confidence intervals. The solid filled region is the 95\% confidence interval determined using pseudo-datasets (Pseud.) and the full signal region. } \label{fig:energygrowth_ttWjttZ}
\end{figure}

It is informative to visualize the benefits of exploiting the energy-growth dependency of the EFT operators using the $\pT(l_0)$ observable. Figure~\ref{fig:energygrowth_ttWjttZ} shows the observed two-dimensional 95\% confidence interval contours, using the asymptotic likelihood approach, as a function of the energy-growth dependency of the EFT operators. Fitting from only the first bin of the signal region, with the lowest \pT, to all of the signal region up to the highest \pT bin, the improvement in the limits is clear from the decreasing size of the contours.
Figure~\ref{fig:energygrowth_ttWjttZ} also shows the complementarity from considering both the \ttWjEW and \ttZ processes together. The limits from the \ttZ process alone, which has larger linear contributions and greater degeneracies between the two EFT operators, is shown by the outermost dotted contour. The degeneracies are reduced by the inclusion of the \ttWjEW process in the fit.

\begin{table}[htbp]
\centering
\caption{Observed one-dimensional and marginalized 95\% confidence intervals on the \cpt and \cHQone Wilson coefficients for different fit configurations using pseudo-datasets.}
\renewcommand{\arraystretch}{1.3}
\begin{tabular}{
l
l
>{\centering\arraybackslash}p{5.2cm}
>{\centering\arraybackslash}p{5.5cm}
}
\toprule
\multirow{2}{*}{Operators} & \multirow{2}{*}{Processes} & \multicolumn{2}{c}{95\% CI (observed) [$\textrm{\TeV}^{-2}$]} \\
& & $\cpt$& $\cHQone$\\
\midrule
\Opt               & \ttWjEW             & $[-10, -2.7]\mathbin{\bigcup}[3.3, 9.6]$   & -- \\
\OHQone            & \ttWjEW             & --                                         & $[-11, -4.5]\mathbin{\bigcup}[4.3, 13]$ \\
\Opt + \OHQone     & \ttWjEW             & $[-9.9, 9.3]$                              & $[-12, 14]$ \\
\midrule
\Opt               & \ttWjEW + \ttZ      & $[-10, 3.2]$                               & -- \\
\OHQone            & \ttWjEW + \ttZ      & --                                         & $[-11, 7.4]$ \\
\Opt + \OHQone     & \ttWjEW + \ttZ      & $[-10, -0.96]\mathbin{\bigcup}[1.5, 7.3]$ & $[-12, 0.33]\mathbin{\bigcup}[3.4, 12]$ \\
\bottomrule
\end{tabular}
\label{tab:all_EFT_limits_95}
\end{table}


%
\FloatBarrier

\section{Conclusions}
\label{sec:Conclusions}

A search is presented for $\ttWjEW$ production in proton--proton collisions using data collected with the ATLAS detector corresponding to an integrated luminosity of 140~$\mathrm{fb^{-1}}$ at $\sqrt{s}=13$~TeV. From the search, limits are set on the Standard Model \ttWjEW process and on effective field theory operators modifying the couplings of the top quark to the $Z$ boson. The search uses same-charge pairs of electrons and muons together with jets, out of which at least one is \btagged, and uses the properties of the most forward jet to discriminate \ttWjEW production from $t\bar{t}W$ QCD background.

The search is affected by the $t\bar{t}W$ and fake-lepton backgrounds. State-of-the-art modeling is used for $t\bar{t}W$ and data-driven techniques are used for the fake-lepton yield estimate, with special attention paid to the interplay between the two.

The observed (expected) normalization factor is $\mu_{\ttWjEW} = 0.9^{+2.4}_{-3.0}$ $(1.0^{+2.4}_{-2.5})$, resulting in an observed 95\% $\textrm{CL}_{s}$ upper limit of $\mu_{\ttWjEW} < 5.3$. The corresponding 95\% $\textrm{CL}_{s}$ upper limit on the observed cross section is $\sigma_{\ttWjEW} < 251$~fb, to be compared with the Standard Model cross section of $47.7^{+1.9}_{-1.4}$ fb~\cite{Frederix:2021agh}. The results from the measurements are consistent with the Standard Model expectations, and are limited by systematic uncertainties.

The unique $tW$-scattering vertex embedded within the \ttWjEW process is exploited to gain new sensitivity to EFT operators.
The interplay between different processes (\ttWjEW and \ttZ) and different operators (\Opt and \OHQone) is investigated, demonstrating the benefits of a combined analysis of this type with a detailed case study.
Considering both processes when looking at \Opt and \OHQone simultaneously closes a degenerate direction present when considering the effects of these operators on \ttZ production alone.
The EFT sensitivity is dominated by uncertainties due to the limited number of events in data, especially at high \pT. Increased sensitivity is expected with the LHC Run 3 data where the interplay between processes with and without sensitivity from significant high energy growth will be even more interesting.


\pagebreak
\FloatBarrier

\section*{Acknowledgments}
%

%

%
%

%
%

We thank CERN for the very successful operation of the LHC and its injectors, as well as the support staff at
CERN and at our institutions worldwide without whom ATLAS could not be operated efficiently.

The crucial computing support from all WLCG partners is acknowledged gratefully, in particular from CERN, the ATLAS Tier-1 facilities at TRIUMF/SFU (Canada), NDGF (Denmark, Norway, Sweden), CC-IN2P3 (France), KIT/GridKA (Germany), INFN-CNAF (Italy), NL-T1 (Netherlands), PIC (Spain), RAL (UK) and BNL (USA), the Tier-2 facilities worldwide and large non-WLCG resource providers. Major contributors of computing resources are listed in Ref.~\cite{ATL-SOFT-PUB-2025-001}.

We gratefully acknowledge the support of ANPCyT, Argentina; YerPhI, Armenia; ARC, Australia; BMWFW and FWF, Austria; ANAS, Azerbaijan; CNPq and FAPESP, Brazil; NSERC, NRC and CFI, Canada; CERN; ANID, Chile; CAS, MOST and NSFC, China; Minciencias, Colombia; MEYS CR, Czech Republic; DNRF and DNSRC, Denmark; IN2P3-CNRS and CEA-DRF/IRFU, France; SRNSFG, Georgia; BMFTR, HGF and MPG, Germany; GSRI, Greece; RGC and Hong Kong SAR, China; ICHEP and Academy of Sciences and Humanities, Israel; INFN, Italy; MEXT and JSPS, Japan; CNRST, Morocco; NWO, Netherlands; RCN, Norway; MNiSW, Poland; FCT, Portugal; MNE/IFA, Romania; MSTDI, Serbia; MSSR, Slovakia; ARIS and MVZI, Slovenia; DSI/NRF, South Africa; MICIU/AEI, Spain; SRC and Wallenberg Foundation, Sweden; SERI, SNSF and Cantons of Bern and Geneva, Switzerland; NSTC, Taipei; TENMAK, T\"urkiye; STFC/UKRI, United Kingdom; DOE and NSF, United States of America.

Individual groups and members have received support from BCKDF, CANARIE, CRC and DRAC, Canada; CERN-CZ, FORTE and PRIMUS, Czech Republic; COST, ERC, ERDF, Horizon 2020, ICSC-NextGenerationEU and Marie Sk{\l}odowska-Curie Actions, European Union; Investissements d'Avenir Labex, Investissements d'Avenir Idex and ANR, France; DFG and AvH Foundation, Germany; Herakleitos, Thales and Aristeia programmes co-financed by EU-ESF and the Greek NSRF, Greece; BSF-NSF and MINERVA, Israel; NCN and NAWA, Poland; La Caixa Banking Foundation, CERCA Programme Generalitat de Catalunya and PROMETEO and GenT Programmes Generalitat Valenciana, Spain; G\"{o}ran Gustafssons Stiftelse, Sweden; The Royal Society and Leverhulme Trust, United Kingdom.

In addition, individual members wish to acknowledge support from CERN: European Organization for Nuclear Research (CERN DOCT); Chile: Agencia Nacional de Investigaci\'on y Desarrollo (FONDECYT 1230812, FONDECYT 1240864, Fondecyt 3240661); China: Chinese Ministry of Science and Technology (MOST-2023YFA1605700, MOST-2023YFA1609300), National Natural Science Foundation of China (NSFC - 12175119, NSFC 12275265); Czech Republic: Czech Science Foundation (GACR - 24-11373S), Ministry of Education Youth and Sports (ERC-CZ-LL2327, FORTE CZ.02.01.01/00/22\_008/0004632), PRIMUS Research Programme (PRIMUS/21/SCI/017); EU: H2020 European Research Council (ERC - 101002463); European Union: European Research Council (BARD No. 101116429, ERC - 948254, ERC 101089007), European Regional Development Fund (SMASH COFUND 101081355, SLO ERDF), European Union, Future Artificial Intelligence Research (FAIR-NextGenerationEU PE00000013), Italian Center for High Performance Computing, Big Data and Quantum Computing (ICSC, NextGenerationEU); France: Agence Nationale de la Recherche (ANR-21-CE31-0022, ANR-22-EDIR-0002); Germany: Deutsche Forschungsgemeinschaft (DFG - 469666862, DFG - CR 312/5-2); China: Research Grants Council (GRF); Italy: Istituto Nazionale di Fisica Nucleare (ICSC, NextGenerationEU), Ministero dell'Università e della Ricerca (NextGenEU 153D23001490006 M4C2.1.1, NextGenEU I53D23000820006 M4C2.1.1, NextGenEU I53D23001490006 M4C2.1.1, SOE2024\_0000023); Japan: Japan Society for the Promotion of Science (JSPS KAKENHI JP22H01227, JSPS KAKENHI JP22H04944, JSPS KAKENHI JP22KK0227, JSPS KAKENHI JP24K23939, JSPS KAKENHI JP24KK0251, JSPS KAKENHI JP25H00650, JSPS KAKENHI JP25H01291, JSPS KAKENHI JP25K01023); Norway: Research Council of Norway (RCN-314472); Poland: Ministry of Science and Higher Education (IDUB AGH, POB8, D4 no 9722), Polish National Science Centre (NCN 2021/42/E/ST2/00350, NCN OPUS 2023/51/B/ST2/02507, NCN UMO-2019/34/E/ST2/00393, UMO-2022/47/O/ST2/00148, UMO-2023/49/B/ST2/04085, UMO-2023/51/B/ST2/00920, UMO-2024/53/N/ST2/00869); Portugal: Foundation for Science and Technology (FCT); Spain: Ministry of Science and Innovation (MCIN \& NextGenEU PCI2022-135018-2, MICIN \& FEDER PID2021-125273NB, RYC2019-028510-I, RYC2020-030254-I, RYC2021-031273-I, RYC2022-038164-I); Sweden: Carl Trygger Foundation (Carl Trygger Foundation CTS 22:2312), Swedish Research Council (Swedish Research Council 2023-04654, VR 2021-03651, VR 2022-03845, VR 2022-04683, VR 2023-03403, VR 2024-05451), Knut and Alice Wallenberg Foundation (KAW 2018.0458, KAW 2022.0358, KAW 2023.0366); Switzerland: Swiss National Science Foundation (SNSF - PCEFP2\_194658); United Kingdom: Royal Society (NIF-R1-231091); United States of America: U.S. Department of Energy (ECA DE-AC02-76SF00515), Neubauer Family Foundation.

%
%


%
%
%
%
%
%

%
%

%
%
%
\printbibliography
\clearpage
 
\begin{flushleft}
\hypersetup{urlcolor=black}
{\Large The ATLAS Collaboration}

\bigskip

\AtlasOrcid[0000-0002-6665-4934]{G.~Aad}$^\textrm{\scriptsize 103}$,
\AtlasOrcid[0000-0001-7616-1554]{E.~Aakvaag}$^\textrm{\scriptsize 17}$,
\AtlasOrcid[0000-0002-5888-2734]{B.~Abbott}$^\textrm{\scriptsize 122}$,
\AtlasOrcid[0000-0002-0287-5869]{S.~Abdelhameed}$^\textrm{\scriptsize 118a}$,
\AtlasOrcid[0000-0002-1002-1652]{K.~Abeling}$^\textrm{\scriptsize 55}$,
\AtlasOrcid[0000-0001-5763-2760]{N.J.~Abicht}$^\textrm{\scriptsize 49}$,
\AtlasOrcid[0000-0002-8496-9294]{S.H.~Abidi}$^\textrm{\scriptsize 30}$,
\AtlasOrcid[0009-0003-6578-220X]{M.~Aboelela}$^\textrm{\scriptsize 45}$,
\AtlasOrcid[0000-0002-9987-2292]{A.~Aboulhorma}$^\textrm{\scriptsize 36e}$,
\AtlasOrcid[0000-0001-5329-6640]{H.~Abramowicz}$^\textrm{\scriptsize 156}$
% ...
\bigskip
\\

% ...
$^{17}$Department for Physics and Technology, University of Bergen, Bergen; Norway.\\
% ...
$^{30}$Physics Department, Brookhaven National Laboratory, Upton NY; United States of America.\\
% ...
$^{36}$$^{(a)}$Facult\'e des Sciences Ain Chock, Universit\'e Hassan II de Casablanca;$^{(b)}$Facult\'{e} des Sciences, Universit\'{e} Ibn-Tofail, K\'{e}nitra;$^{(c)}$Facult\'e des Sciences Semlalia, Universit\'e Cadi Ayyad, LPHEA-Marrakech;$^{(d)}$LPMR, Facult\'e des Sciences, Universit\'e Mohamed Premier, Oujda;$^{(e)}$Facult\'e des sciences, Universit\'e Mohammed V, Rabat;$^{(f)}$Institute of Applied Physics, Mohammed VI Polytechnic University, Ben Guerir; Morocco.\\
% ...
$^{45}$Physics Department, Southern Methodist University, Dallas TX; United States of America.\\
% ...
$^{49}$Fakult\"{a}t Physik , Technische Universit{\"a}t Dortmund, Dortmund; Germany.\\
% ...
$^{55}$II. Physikalisches Institut, Georg-August-Universit\"{a}t G\"ottingen, G\"ottingen; Germany.\\
% ...
$^{103}$CPPM, Aix-Marseille Universit\'e, CNRS/IN2P3, Marseille; France.\\
% ...
$^{118}$$^{(a)}$New York University Abu Dhabi, Abu Dhabi;$^{(b)}$United Arab Emirates University, Al Ain; United Arab Emirates.\\
% ...
$^{122}$Homer L. Dodge Department of Physics and Astronomy, University of Oklahoma, Norman OK; United States of America.\\
% ...
$^{156}$Raymond and Beverly Sackler School of Physics and Astronomy, Tel Aviv University, Tel Aviv; Israel.\\
% ...
$^{a}$ Also at Affiliated with an institute formerly covered by a cooperation agreement with CERN.\\
$^{b}$ Also at An-Najah National University, Nablus; Palestine.\\
$^{c}$ Also at Borough of Manhattan Community College, City University of New York, New York NY; United States of America.\\
$^{d}$ Also at Center for Interdisciplinary Research and Innovation (CIRI-AUTH), Thessaloniki; Greece.\\
$^{e}$ Also at Centre of Physics of the Universities of Minho and Porto (CF-UM-UP); Portugal.\\
$^{f}$ Also at CERN, Geneva; Switzerland.\\
% ...

\end{flushleft}


%

%
%

%

%
%
%
%

\end{document}